\documentclass[11pt]{article}
\usepackage{epsfig,subfig}
\usepackage{amsmath,amssymb,amscd,amsfonts}
\usepackage{authblk}
\begin{document}

\title{The $Ca^{2+}$-activated $Cl^-$ current ensures robust and reliable signal amplification in vertebrate olfactory receptor neurons}

\author[1]{Johannes Reisert}
\author[2,3]{J\"urgen Reingruber}
\affil[1]{Monell Chemical Senses Center, Philadelphia, PA, USA}
\affil[2]{Ecole Normale Sup\'erieure, 46 rue d'Ulm, 75005 Paris , France.}
\affil[3]{INSERM U1024, Paris, France.}

\date{\today}
\maketitle

\begin{abstract}
Activation of most primary sensory neurons results in transduction currents that are carried by cations. One notable exception is the vertebrate olfactory receptor neuron (ORN), where the transduction current is carried largely by the anion $Cl^-$. However, it remains unclear why ORNs use an anionic current for signal amplification. We have sought to provide clarification on this topic by studying the so far neglected dynamics of $Na^+$, $Ca^{2+}$, $K^+$ and $Cl^-$ in the small space of olfactory cilia during an odorant response. Using computational modeling and simulations we compared the outcomes of signal amplification based on either $Cl^-$ or $Na^+$ currents. We found that amplification produced by $Na^+$ influx instead of a $Cl^-$ efflux is problematic due to several reasons:  First, the $Na^+$ current amplitude varies greatly depending on mucosal ion concentration changes. Second, a $Na^+$ current leads to a large increase in the ciliary $Na^+$ concentration during an odorant response. This increase inhibits and even reverses $Ca^{2+}$ clearance by $Na^+/Ca^{2+}/K^+$ exchange, which is essential for response termination. Finally, a $Na^+$ current increases the ciliary osmotic pressure, which could cause swelling to damage the cilia. By contrast, a transduction pathway based on $Cl^-$ efflux circumvents these problems and renders the odorant response robust and reliable.    
\end{abstract}

\section{Introduction }
\begin{figure}%[tbhp]
 	\centering
 	\includegraphics[width=0.9\linewidth]{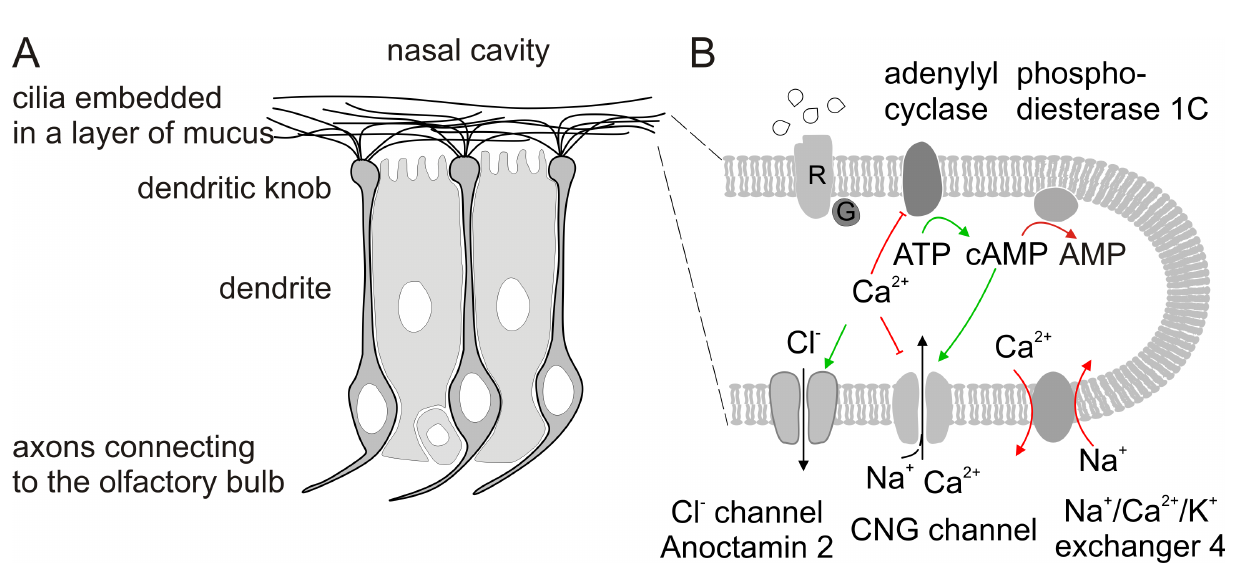}
 	\caption{(A) Schematic of the olfactory epithelium with olfactory receptor neurons. (B) The magnification shows the transduction pathway in a cilium. Arrows in green and red show events for response activation and termination. R is odorant receptor and G is G protein. Modified from \cite{ReisertZhao_JGP2011}.}  
 	\label{FigIntroduction} 
 \end{figure}

Olfactory receptor neurons (ORNs) in the nasal olfactory epithelium are the fundamental neurons underlying the sense of smell \cite{Kleene_Review2008}. ORNs are bipolar neurons that extend an axon to the olfactory bulb and a single dendrite to the epithelial border, ending in the dendritic knob (Fig.~\ref{FigIntroduction}A). The cellular compartments that detect and transduce odorants into an electrical signal are the olfactory cilia. There are about 10-15 long and slender cilia per ORN embedded in the mucus on the surface of the olfactory epithelium. The ciliary membrane contains odorant receptors that are activated by odorant molecules carried into the nose during inhalation and dissolved in the mucus. Receptor activation initiates a biochemical transduction cascade that depolarizes the neuron via the opening of ion channels located in the ciliary membrane (Fig.~\ref{FigIntroduction}B).

Unlike in other sensory systems like taste, hearing or phototransduction \cite{BookFainSensoryTransduction}, a remarkable feature of olfactory signal transduction is that it involves not only the opening of cation channels but also of anion channels (Fig.~\ref{FigIntroduction}B). Receptor activation first leads, via a G protein, to activation of adenylyl cyclase 3 (AC3) that synthesizes cAMP. The subsequent increase in the ciliary cAMP concentration opens cAMP-gated, cyclic nucleotide-gated (CNG) cation channels that are permeable to $Ca^{2+}$ and to a lesser extent to $Na^+$ and $K^+$ \cite{NakamuraGold_CNG_Nature1987,FiresteinZufallShepherd_CNG_JNsc1991}. The $Ca^{2+}$ influx triggers the opening of an secondary anionic channel: $Ca^{2+}$-gated $Cl^-$ channel (Anoctamin 2, Ano2, also known as TMEM16B) \cite{PifferiEtal_2009,RascheEtal_2010,KleeneeGestland_ChlorideConductanceFrog_JNsc1991,StephanReisertetal_Ano2_PNAS2009}. $Cl^-$ efflux ensues, as intracellular $Cl^-$ is high in ORNs \cite{Reisertetal_Nkcc1_Neuron2005,KanekoEtal_ClAccumulation_JNsc2004}. Ciliary $Ca^{2+}$ is predominantly removed by electrogenic $Na^+/K^+/Ca^{2+}$ exchange (NCKX4) \cite{StephanReisertetal_NCKX4_NatNeurosc2011}. The activation of the Ano2 channels is the main amplification step, and most of the transduction current is carried by $Cl^-$ efflux \cite{LiBenChaimYauLin_PNAS2016,Billigetal_Ano2_JPhys2011,Reisertetal_ClCurrent_JGenPhys2003,LoweGold_Noise_Nature1993,Kleene_Neuron1993}. This large amplification is thought to be the principal function of the $Cl^-$ current. When the $Cl^-$ current is deleted either pharmacologically or by deletion of the Ano2 gene, mice retain some sense of smell due to the CNG channels, but ORN responses are much smaller affecting sensitivity at signaling threshold, the ability to track novel odorants, and spiking activity  \cite{Billigetal_Ano2_JPhys2011,LiYau_PNAS2018,NeureitherEtal_2017,PietraEtal_2016}.
 
It is still unclear why ORNs rely upon a $Cl^-$ current to boost the amplitude of their signal \cite{DibattistaEtal_Review2017,Frings_PNAS2016}. A larger signal could be produced if ORNs had more CNG channels, but a simple increase in channel number would produce larger influx of both $Na^+$ and $Ca^{2+}$, which would disturb $Ca^{2+}$ homoeostasis and affect several $Ca^{2+}$ dependent feedback processes in the biochemical transduction pathway \cite{Kleene_Review2008,Reisert_Review2005}. Moreover, a cilium has as volume of only about 0.5 femtoliter, and a current of just 1 pA already produces an ion flux of about 20 mM per second. A large $Na^+$ influx would greatly increase the intracellular $Na^+$ concentration with currently unclear consequences. These problems could conceivably be obviated by the use of a current based on $Cl^-$ efflux. A $Cl^-$ current might also reduce response variation caused by changes in mucosal ion concentrations \cite{KurahashiYau_CurrBiol1994,KurahashiYau_ClCurrent_Nature1993}. The cilia of ORNs must function reliably while embedded in mucus exposed to the outside world. Mucosal ion concentrations can be altered by exposure to water or sneezing, as well as by stimulation of the parasympathetic and $\beta$-adrenergic systems in mice or human patients with airway disease (e.g. cystic fibrosis) \cite{KozlovaEtal_2006}. 

Since no experimental method presently allows for the measurement of ion concentrations in the small volume of a cilium, we used mathematical modeling and response simulation to address the role of the $Cl^-$ current. We developed a model of olfactory transduction that not only incorporates the biochemical transduction pathway but also includes spatial resolution of ion dynamics in the volume of the cilium during an odorant response. To study the impact of the current carrier, we modeled and compared signal amplification for two complementary scenarios: a biological scenario where amplification is based on a $Cl^-$ current, and an artificial scenario where amplification is based on a $Na^+$ current. One possibility to model the artificial scenario would have been to remove the Ano2 channels and increase the $Na^+$ current by simply augmenting the number of CNG channels. However, this would have required additional substantial changes in the transduction part of the model to adjust the cAMP dynamics such that odorant responses with a cAMP-activated $Na^+$ current become comparable to the biological responses with a $Ca^{2+}$-activated $Cl^-$ current. We therefore have elected to keep the CNG channels unchanged and to postulate a $Ca^{2+}$-activated $Na^+$ current, similar to the $Ca^{2+}$-activated $Cl^-$ current. In this way, we could keep all aspects of the transduction pathway unchanged between the two scenarios, and we could neatly compare the effects of amplification based on $Na^+$ influx and $Cl^-$ efflux without complicating factors. Our general conclusions apply however equally well to any increase in $Na^+$ influx regardless of the mechanism of activation of the $Na^+$ current. We show that signal amplification based on $Na^+$ influx can produce responses comparable to those produced by $Cl^-$ efflux, but a mechanism based on $Na^+$ renders the response much less stable to changes in mucosal ion concentrations. Moreover, a $Na^+$ current leads to a large increase in intracellular $Na^+$, which inhibits and can even reverse $Ca^{2+}$ extrusion by $Na^+/Ca^{2+}/K^+$ exchange. Finally, a $Na^+$ current leads to an increase in the ciliary osmotic pressure, which could conceivably produce swelling that damages the cilia. All of these detrimental consequences are avoided by the use of a $Cl^-$ current, which not only amplifies the response but also keeps it robust and reliable.   

 \begin{figure}%[tbhp]
 	\centering
 	\includegraphics[width=0.9\linewidth]{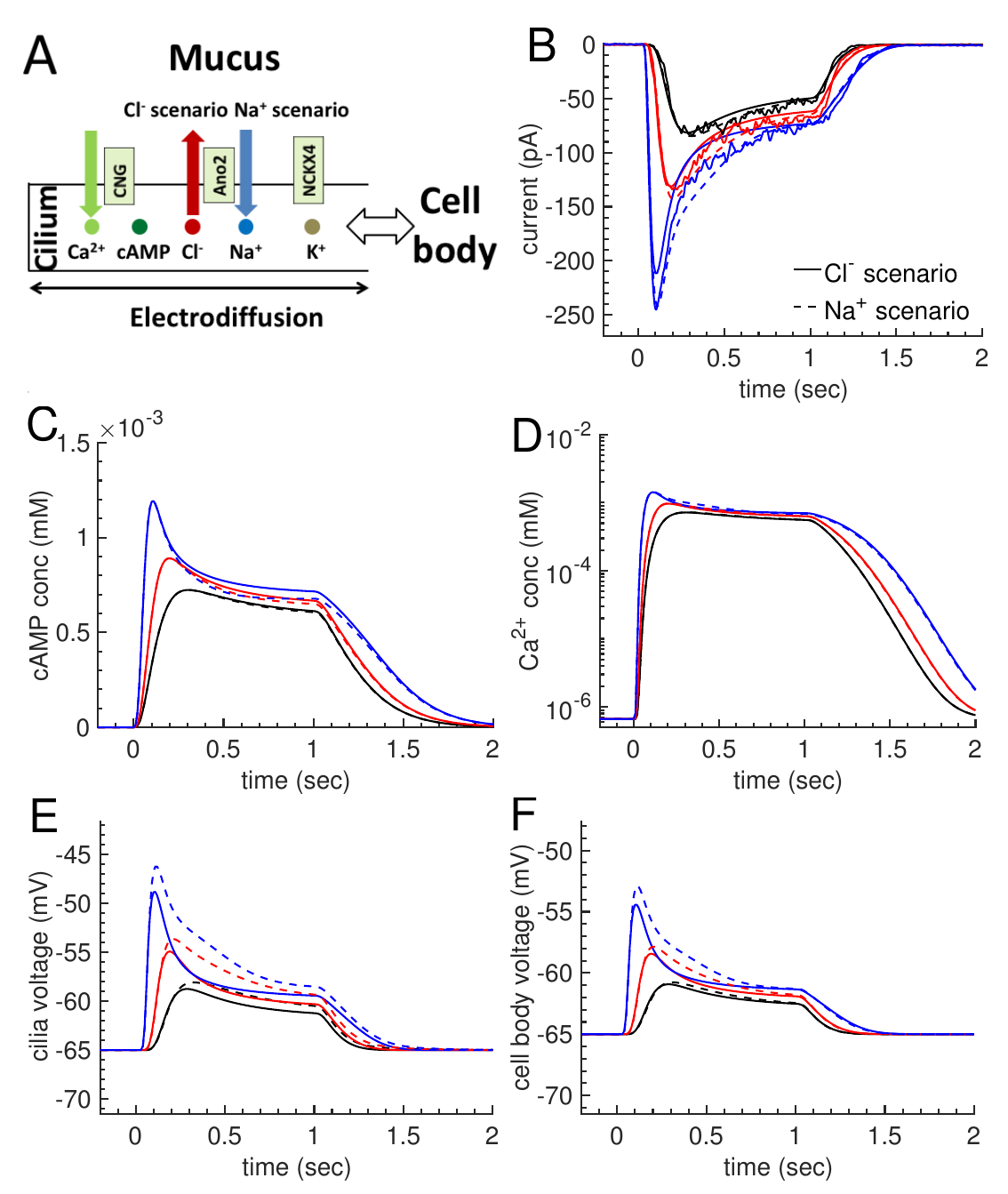}
 	\caption{(A) Schematic of the ciliary electrodiffusion model with the two scenarios having $Cl^-$ or $Na^+$ as carrier of the transduction current. (B) Comparison of dorant-induced responses (noisy traces) from a mouse olfactory receptor neuron with simulation results for three different one-second odorant stimulations of increasing concentrations (10, 30 and 100 $\mu$M). Continuous lines correspond to the $Cl^-$ and dashed lines to the $Na^+$ scenario. Panels (C-E):  Comparison of results averaged along a cilium for cAMP and $Ca^{2+}$ concentrations and cilium voltage. (F) Cell body depolarization. }  
 	\label{FigSimAveraged} 
 \end{figure}
 
%%%%%%%%%%%%%%%%%%%%%%%%%%%%%%%%%%%%%%%%%%%%%%%%%%%%%%%%
\section{Results }
%%%%%%%%%%%%%%%%%%%%%%%%%%%%%%%%%%%%%%%%%%%%%%%%%%%%%%%%
We compare two complementary scenarios: in the biological scenario (in the following referred to as the $Cl^-$ scenario) amplification is due to $Cl^-$ efflux, whereas in the artificial scenario (referred to as the $Na^+$ scenario) amplification is due to $Na^+$ influx (Fig.~\ref{FigSimAveraged}A). To minimize the differences between the two scenarios, in the $Na^+$ scenario we introduce artificial Ano2 channels that have the same $Ca^{2+}$ activation properties as the biological Ano2 channels but are permeable to $Na^+$ instead of $Cl^-$. For simplicity, we also refer to them as Ano2 channels, since from the scenario it becomes clear which type is considered. Hence, both scenarios are absolutely identical except for the Ano2 permeability. We further simplify and neglect the $Na^+$ and $K^+$ currents through CNG channels because they are small compared to the Ano2 current \cite{Reisertetal_ClCurrent_JGenPhys2003,Billigetal_Ano2_JPhys2011,DibattistaEtal_Review2017} and therefore only have a minor impact on ion dynamics. Moreover, in this way both scenarios are clearly distinguished, and differences can be unambiguously attributed to effects evoked by the two current carriers.

%%%%%%%%%%%%%%%%%%%%%%%%%%%%%%%%%%%%%%%%%%%%%%%%%%%%%%%%
\subsection{A $Na^+$ current induces large $Na^+$ fluctuations in a cilium}
%%%%%%%%%%%%%%%%%%%%%%%%%%%%%%%%%%%%%%%%%%%%%%%%%%%%%%%%
We calibrated our model by fitting suction-electrode recordings from a mouse ORN in response to a one-second stimulation with three different odorant concentrations (for details see the SI). In Fig.~\ref{FigSimAveraged}B we compare the computed current between mucus and 15 identical cilia with the experimental data. The simulations for both scenarios are performed with exactly the same set of parameters; only the Ano2 permeability in each scenario is different. 

In Fig.~\ref{FigSimAveraged}C-F, we compare results that are spatially averaged along the length of a cilium for quantities that are commonly used in standard olfactory models: cAMP and $Ca^{2+}$ concentration, ciliary and cell body depolarization. Fig.~\ref{FigSimAveraged} does not suggest any advantage for $Cl^-$ as current carrier since simulation results are very similar in the two scenarios. More generally, with a standard cable theory model where ionic concentration changes and their effects are neglected (except for $Ca^{2+}$), it makes no difference for the electrical response whether the current is carried by $Cl^-$ or $Na^+$. With such a model the simulations in Fig.~\ref{FigSimAveraged} for both scenarios would exactly match. 

\begin{figure}%[tbhp]
	\centering
	\includegraphics[width=0.9\linewidth]{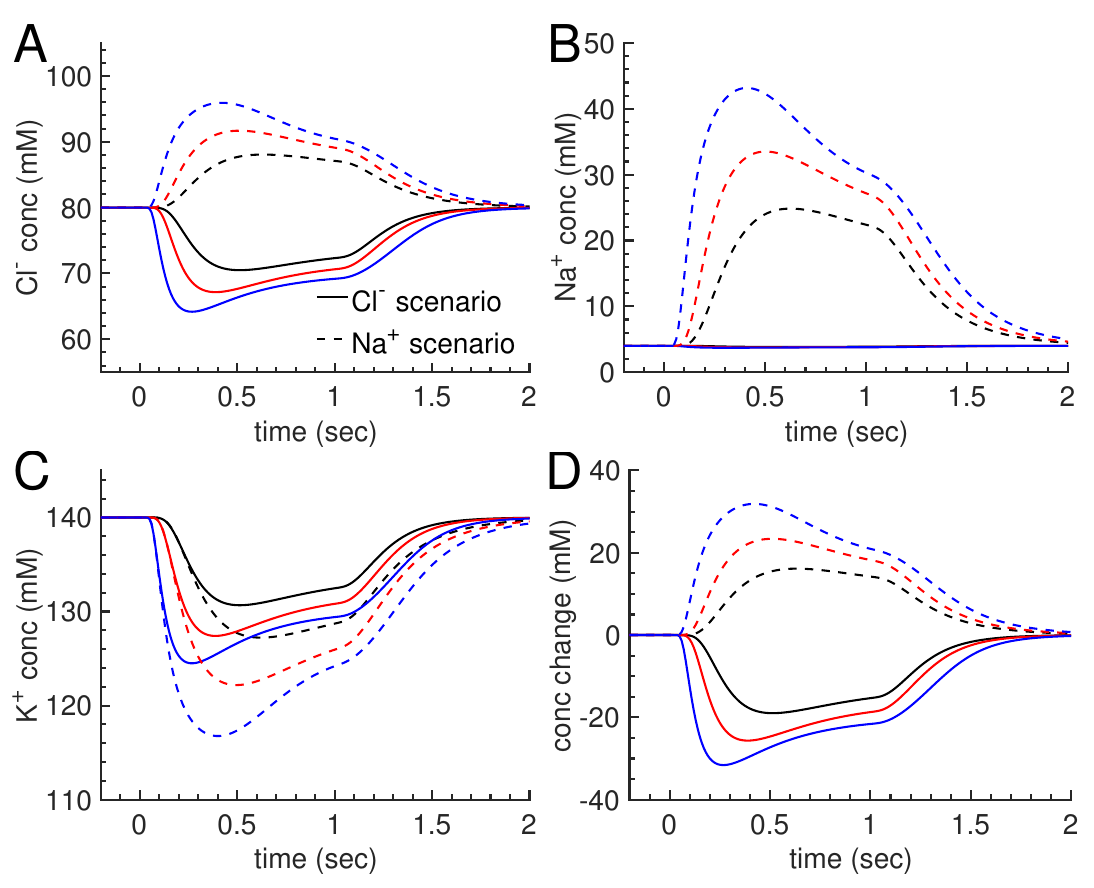}
	\caption{Spatially averaged ion concentrations in a cilium.  (A-C) Comparison of the dynamics for $Cl^-$, $Na^+$ and $K^+$ concentrations corresponding to the odorant responses from Fig.~\ref{FigSimAveraged}. (D) Osmotic concentration change.} 
	\label{FigIonConcAveraged} 
\end{figure}

The spatially averaged results for the $Na^+$, $Cl^-$ and $K^+$ concentrations, on the other hand, reveal large differences between the two scenarios (Fig.~\ref{FigIonConcAveraged}). In the $Cl^-$ scenario, the ciliary $Cl^-$ concentration drops due to $Cl^-$ efflux into the mucus (Fig.~\ref{FigIonConcAveraged}A). In strong contrast, the $Na^+$ concentration changes very little (Fig.~\ref{FigIonConcAveraged}B). Note that we neglect $Na^+$ influx through CNG channels, which would lead to a small increase in the ciliary $Na^+$ concentration instead of the slight decrease seen in Fig.~\ref{FigIonConcAveraged}B. The $K^+$ concentration decreases due to efflux into the cell body driven by the ciliary depolarization (Fig.~\ref{FigIonConcAveraged}C). $Na^+$ is less affected because the internal $Na^+$ concentration is much lower (4 mM). In the $Cl^-$ scenario the decline of $K^+$ and $Cl^-$ electrically compensate one another such that electroneutrality is preserved. The situation changes with a $Na^+$ current. With $Na^+$ influx the ciliary $Na^+$ concentration is dramatically increased by up to 40 mM (Fig.~\ref{FigIonConcAveraged}B). The ciliary depolarization causes the $Cl^-$ concentration to increase (Fig.~\ref{FigIonConcAveraged}A) and the $K^+$ concentration to decrease (Fig.~\ref{FigIonConcAveraged}C) via exchanges with the cell body. The changes in the $K^+$ and $Cl^-$ concentrations no longer electrically compensate as in the $Cl^-$ scenario, but instead they generate a large gap of positive charge that is filled with $Na^+$ flowing in from the mucus to maintain electroneutrality. Finally, we note that the overall ion concentration in a cilium drops with a $Cl^-$ current, whereas it increases with a $Na^+$ current (Fig.~\ref{FigIonConcAveraged}D). Hence, osmotic concentration and pressure change in opposite ways in the two scenarios. In summary, we found complex and highly coupled ion dynamics in the cilia due to the small ciliary volume and the requirement for electroneutrality (see also Figs. S2 and S3 in the SI). 

%%%%%%%%%%%%%%%%%%%%%%%%%%%%%%%%%%%%%%%%%%%%%%%%%%%%%%%%
\subsection{Large ionic concentration gradients are generated in a cilium during an odorant response}
\begin{figure}%[tbhp]
	\centering
	\includegraphics[width=0.9\linewidth]{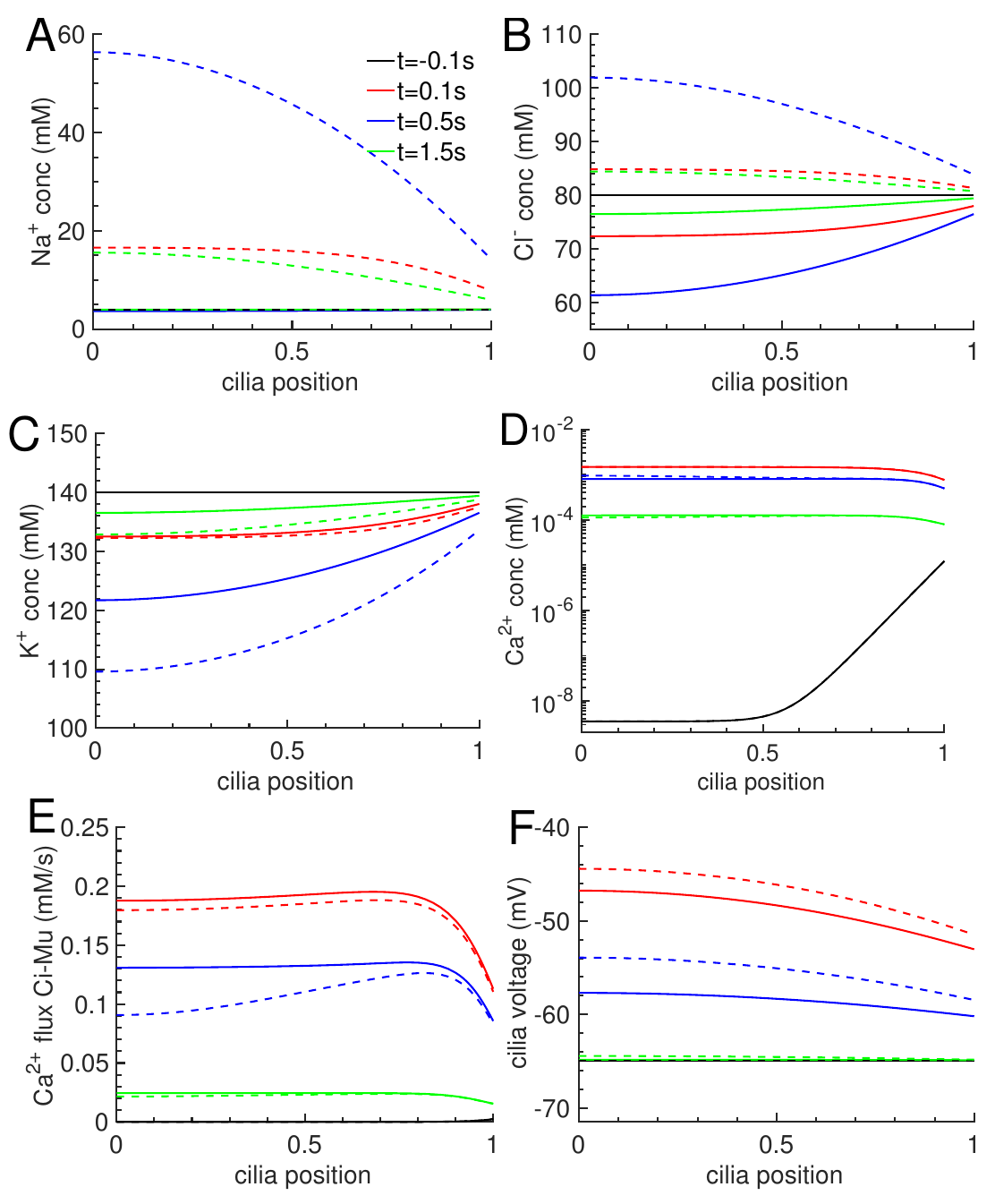}
	\caption{Spatial distributions along a cilium at four different times. The rescaled ciliary position is between zero and one. Zero marks the tip and one the base of a cilium. (A-D) show $Na^{+}$, $Cl^{-}$, $K^{+}$ and $Ca^{2+}$ gradients, (E) $Ca^{2+}$ efflux due to NCKX4 exchange and (F) the ciliary voltage. Continuous lines show result from the $Cl^-$ scenario, dashed lines from the $Na^+$ scenario. The distributions correspond to Fig.~\ref{FigSimAveraged} and Fig.~\ref{FigIonConcAveraged} for the strongest odorant stimulation (100 $\mu M$).  }  
	\label{FigIonConcSpatial} 
\end{figure}

%%%%%%%%%%%%%%%%%%%%%%%%%%%%%%%%%%%%%%%%%%%%%%%%%%%%%%%%
Next, we investigated the degree of spatial inhomogeneity that is generated in a cilium during an odorant response. In Fig.~\ref{FigIonConcSpatial} we compare spatial distributions for ion concentrations, $Ca^{2+}$ efflux via the exchangers, and ciliary voltage for the strongest odorant stimulation in Fig.~\ref{FigSimAveraged} at four different times: before, twice during and once after the one-second odorant stimulus. Continuous lines correspond to the $Cl^-$ scenario, dashed lines to the $Na^+$ scenario. In agreement with Fig.~\ref{FigIonConcAveraged}, we find large differences between the two scenarios for the $Cl^-$ and $Na^+$ concentrations (Fig.~\ref{FigIonConcSpatial}A,B), and to a lesser extent for the $K^+$ concentration (Fig.~\ref{FigIonConcSpatial}C). With $Cl^-$ as the current carrier, almost no $Na^+$ gradient is generated during an odorant response. This is in strong contrast to a $Na^+$ current that generates large $Na^+$ gradients. At the tip of a cilium, $Na^+$ influx pushes the $Na^+$ concentration to values that are almost 15 fold larger compared to the basal value of 4 mM (Fig.~\ref{FigIonConcSpatial}A, blue dashed line). As a consequence, the NCKX4 exchangers are inhibited, leading to reduced $Ca^{2+}$ efflux along the cilium in the $Na^+$ scenario (Fig.~\ref{FigIonConcSpatial}E). The $Cl^-$ gradients have opposite behaviors in the two scenarios (Fig.~\ref{FigIonConcSpatial}B). In the $Cl^-$ scenario, $Cl^-$ efflux leads to a distribution where the concentration is minimal near the ciliary tip. In contrast, a $Na^+$ current leads to a $Cl^-$ increase that is maximal near the tip as $Cl^-$ flows into the cilia from the cell body due to the ciliary depolarization (Fig.~\ref{FigIonConcSpatial}F). The $K^+$ concentration decreases towards the tip of a cilium, but the $K^+$ gradients are qualitatively similar between the two scenarios; however, a $Na^+$ current entails much higher $K^+$ gradients (Fig.~\ref{FigIonConcSpatial}C). In contrast to the strong gradients generated for $Cl^-$, $Na^+$ and $K^+$, the $Ca^{2+}$ concentration remains rather homogeneous along a cilium during the odorant response (Fig.~\ref{FigIonConcSpatial}D). This is because we assumed that a cilium is uniformly activated by the odorant application, and NCKX4 exchangers that remove $Ca^{2+}$ are uniformly distributed along a cilium. Because we did not include spontaneous receptor activity and a basal cAMP synthesis rate, the resting cAMP concentration vanishes and there is no $Ca^{2+}$ influx via CNG channels in the absence of odorant stimulation. At rest the exchangers therefore reduce the ciliary $Ca^{2+}$ concentration much below nM level (see also the SI). The $Ca^{2+}$ gradient at $t = -0.1 s$ in Fig.~\ref{FigIonConcSpatial}D (black trace) is caused by influx from the cell body where the concentration is 40 nM. Note that at $t=1.5s$ (green trace), the $Ca^{2+}$ concentration did not yet regain its resting distribution. The subtle discrepancies in the $Ca^{2+}$ dynamics between the two scenarios are generated by inhibition of NCKX4 exchange in the $Na^+$ scenario (Fig.~\ref{FigIonConcSpatial}E). Ultimately, this is the reason for the differences in the Ano2 currents between the two scenarios in Fig.~\ref{FigSimAveraged}A. One might be surprised that the generated $Ca^{2+}$ discrepancies are not larger. However, the parameters are specifically fitted such that the currents in both scenarios agree with the experimental data. Since the currents reflect the Ano2 open probability that depends very sensitively on $Ca^{2+}$ with a Hill exponent of 2.3 \cite{StephanReisertetal_Ano2_PNAS2009}), the $Ca^{2+}$ concentrations necessarily have to be very similar between the two scenarios. As we will show below, this is no longer the case when we modify parameters, for example by changing mucosal ion concentrations.  

%%%%%%%%%%%%%%%%%%%%%%%%%%%%%%%%%%%%%%%%%%%%%%%%%%%%%%%%
\subsection{With a $Na^+$ current the odorant response is not robust against ionic concentration changes in the mucus}
%%%%%%%%%%%%%%%%%%%%%%%%%%%%%%%%%%%%%%%%%%%%%%%%%%%%%%%%
\begin{figure}%[tbhp]
	\centering
	\includegraphics[width=0.9\linewidth]{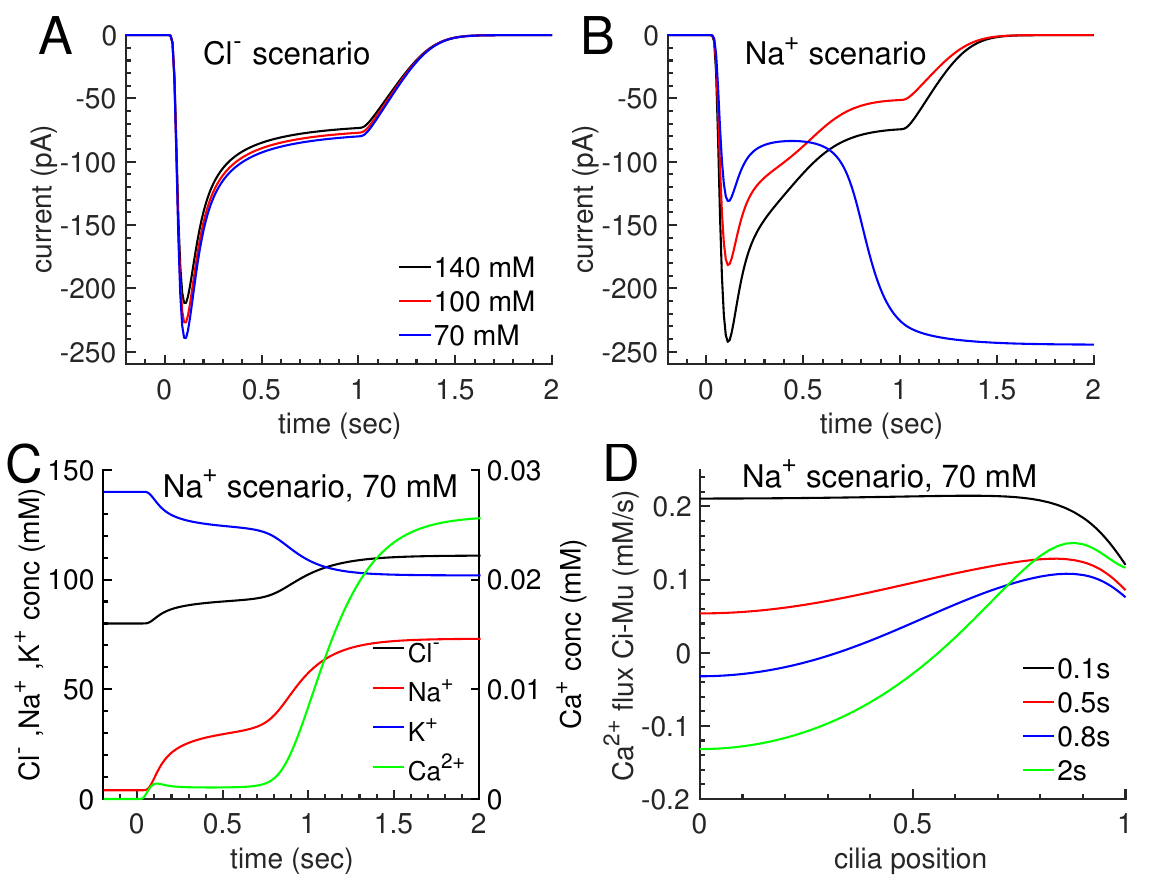}
	\caption{Stability of the odorant response in a varying mucosal environment. (A-B) Comparison of the current for strongest odorant stimulation and mucosal concentrations of $Cl^-$ and $Na^+$ of 140 mM, 100 mM and 70 mM. (A) $Cl-$ scenario. (B) $Na^+$ scenario. (C-D) Results for the $Na^+$ scenario with 70 mM. (C) Spatially averaged ion concentrations. (D) $Ca^{2+}$ flux between cilium and mucus generated by NCKX4 exchange. A positive flux corresponds to efflux from a cilium. }  
	\label{FigMucus} 
\end{figure}

Next, we addressed the important question whether the odorant response is robust against mucosal ion concentration changes. To test robustness, we performed simulations for the highest odorant stimulation with the same parameters as before, but we reduced the mucosal concentrations of $Na^+$ and $Cl^-$ from 140 mM to 100 mM and 70 mM each (Fig.~\ref{FigMucus}A-B). Such values are not unrealistic as several reports suggest that mucosal concentrations can be significantly lower than the 140 mM in Ringer solution used in electrophysiological experiments \cite{ReuterEtal_1998,ChiuEtal_1989,JoshiEtal_1987}. With a $Cl^-$ current, the response is little affected by changing mucosal concentrations (Fig.~\ref{FigMucus}A). Fo example, the peak amplitude of around 240 pA for 70 mM is only slightly increased compared to around 210 pA for 140 mM due to the larger driving force. In contrast, with a $Na^+$ current, the response is strongly altered (Fig.~\ref{FigMucus}B). The initial peak amplitude for 70 mM is almost halved compared to 140 mM. In brief (for a more detailed analysis we refer to Eqs. 12 and 13 in the SI), $Cl^-$ efflux is determined by the ciliary $Cl^-$ concentration; altering the mucosal concentration affects this result only a little. In contrast, $Na^+$ influx is proportional to the mucosal $Na^+$ concentration and is strongly affected by mucosal concentration changes. This result shows that a $Cl^-$ but not a $Na^+$ current is robust against mucosal ion concentration changes. 

Finally, a very surprising effect occurs in the $Na^+$ scenario due to our artificial $Na^+$ channels that are $Ca^{2+}$ activated. When the mucosal concentrations are reduced to 70 mM, at a time around 0.7s, instead of declining, the current increases again and reaches a plateau value that persists even after the odorant stimulation ends at 1s (Fig.~\ref{FigMucus}B, blue line). Correspondingly, intracellular ion concentrations attain plateau values that are very different from their initial values (Fig.~\ref{FigMucus}C). The system is obviously bistable with two stable fixed points in the absence of odorant stimulation: one with zero current and one with a large current. The reason for this bistability is a positive feedback loop that is present if the $Na^+$ channels are $Ca^{2+}$-activated: a sufficiently large $Na^+$ current increases the ciliary $Na^+$ concentration such that the exchangers eventually switch to reverse mode in the frontal part of the cilium (Fig.~\ref{FigMucus}D, negative current near the tip). In reverse mode, the exchangers import $Ca^{2+}$ from the mucus, which keeps the $Ca^{2+}$-activated $Na^+$ channels open even without odorant stimulation. In Fig.~\ref{FigMucus}B, the odorant stimulation pushes the system into the basin of attraction of the non-zero fixed point, which abolishes response termination. In contrast, bistability does not occur with a $Ca^{2+}$-activated $Cl^-$ current because the ciliary $Na^+$ concentration remains low and exchangers do not switch to reverse mode.  

%%%%%%%%%%%%%%%%%%%%%%%%%%%%%%%%%%%%%%%%%%%%%%%%%%%%%%%%
\section{Discussion}
%%%%%%%%%%%%%%%%%%%%%%%%%%%%%%%%%%%%%%%%%%%%%%%%%%%%%%%%
We studied the ciliary ion dynamics during an odorant response to elucidate a long-standing question in olfactory transduction: Why is amplification in olfactory transduction carried by a $Cl^-$ and not a $Na^+$ current. To answer this question, we developed a mathematical model that predicts the spatio-temporal ion dynamics in a cilium. Compared to previous models that focus on the biochemical signal transduction pathway and on the $Ca^{2+}$ dynamics \cite{DePaloetal_BJ2012,Reidletal_CaOsc_BJ2006,Doughertyetal_PNAS2005,SuzukiEtal_OscCurrent_ChemSensesJ2002}, we consider the coupled dynamics between $Cl^-$, $Na^+$, $K^+$ and $Ca^{2+}$ ions in the restrained volume of a cilium during an odorant response. Moreover, we did not rely on an effective equation for $Ca^{2+}$ extrusion that depends only on the ciliary $Ca^{2+}$ concentration, but we considered that the NCKX4 exchange is electrogenic and depends on the $Na^+$ and $K^+$ gradients between cilia and mucus. We compared simulations for two complementary scenarios: a biological scenario where signal amplification is based on $Cl^-$ efflux, and an artificial scenario where it is based on $Na^+$ influx. We found significant differences between both scenarios, which offer explanations for why a $Cl^-$ current is preferential for signal amplification in ORNs.

It has been suggested that a $Cl^-$ current renders the odorant response robust against mucosal ion concentration changes \cite{KurahashiYau_CurrBiol1994,KurahashiYau_ClCurrent_Nature1993}. Indeed, our simulations directly show that a current carried by $Cl^-$ efflux is largely insensitive to $Cl^-$ and $Na^+$ concentration changes in the mucus, contrary to a current carried by $Na^+$ influx.

Our work revealed complex and highly coupled ion dynamics in a cilium. We found striking differences between the two scenarios in particularly for the dynamics of the $Cl^-$ and $Na^+$ concentrations. With $Cl^-$ efflux, the ciliary $Cl^-$ concentration decreases, as expected, but surprisingly the $Na^+$ concentration remains largely unaffected. In contrast, with $Na^+$ influx, both $Cl^-$ and $Na^+$ concentrations significantly increase during an odorant response. This has two major consequences: First, the large $Na^+$ concentration inhibits $Ca^{2+}$ clearance by NCKX4 exchange, which affects response termination and other transduction processes that depend on $Ca^{2+}$ feedback \cite{Kleene_Review2008}. This is not the case in the $Cl^-$ scenario where the $Na^+$ concentration remains low and stable and fast $Ca^{2}$ clearance is assured. Second, whereas the osmotic concentration in a cilium decreases with a $Cl^-$ current, it increases with a $Na^+$ current. This might be of relevance because an increase in osmotic concentration necessarily entails water influx that might compromise the stability of the fragile cilia. 

In line with steady state results obtained previously by Lindemann \cite{Lindemann_BJ2001}, we found that large ciliary concentration gradients for $Cl^-$ and $K^+$ are generated in both scenario. Unexpectedly, with a $Cl^-$ current almost no $Na^+$ gradient is generated, which ensures homogeneous $Ca^{2+}$ clearance along the whole cilium. In contrast, a large $Na^+$ gradient is generated in the $Na^+$ scenario, which inhibits and can even reverse exchanger function in the frontal part of a cilium. 

These differences between both scenarios are generic and a consequence of the current carrier (but see next paragraph). They occur independently of the precise type of $Cl^-$ or $Na^+$ channels that are involved. Thus, signal amplification based on increasing the number of CNG channels would suffer from the same problems generated by a large $Na^+$ influx. Moreover, since CNG channels are permeable to $Ca^{2+}$ and $Na^+$, increasing their number would increase $Ca^{2+}$ influx and at the same time reduce $Ca^{2+}$ clearance by NCKX4 exchange due to increased intracellular $Na^+$. This has the potential to strongly alter the $Ca^{2+}$ dynamics. Robust $Ca^{2+}$ homoeostasis is however essential because of multiple $Ca^{2+}$-dependent feedback process in the biochemical transduction pathway. 

We found a striking effect that is specific to the fact that we introduced an artificial $Na^+$ channel that is $Ca^{2+}$-activated. With such a channel type we found a bistable system when we reduced the mucosal $Na^+$ concentration. In this case, the reduced $Na^+$ driving force together with ciliary depolarization, decreasing $K^+$ and increasing $Na^+$ concentrations inhibited and reversed the exchanger mode in the frontal part of a cilium at a strong stimulation. Part of the exchangers now import $Ca^{2+}$ from the mucus, which keeps the $Ca^{2+}$-activated $Na^+$ channels open and the current flowing even after the odorant stimulation is over. Thus, an olfactory pathway based on $Ca^{2+}$-activated $Na^+$ channels would pose a serious threat to robustness and reliability of odorant detection. Interestingly, although $Ca^{2+}$-activated $K^+$ channels are ubiquitous and important to control neuronal excitability and many other physiological processes \cite{LatorreEtal_CaActivPotCh_1989,VergaraEtal_CaActivPotCh_1998}, $Ca^{2+}$-activated $Na^+$ channels seem much less prevalent. We found only very few reports where such channels produce long-lasting action potentials in starfish oocyte or the egg of a nemertean worm \cite{JaffeEtal_CaActivNaCh_1986}. Since $Ca^{2+}$ clearance is often accomplished by $Na^+$-dependent exchangers, the rare occurrence of such channels might be related to the bistable behavior observed here. 

Because of the large number of parameters and non-linear interactions, it is beyond the scope of this work to present a comprehensive analysis and discussion of the parameter space. This will be done in more theoretical future work. Our goal here was to outline the main differences between the $Cl^-$ and $Na^+$ scenarios, and these differences are robust against reasonable changes in parameter values. 

We developed a detailed spatio-temporal model that we applied to analyze ion dynamics in a cilium. However, our model can also be applied to study other puzzling aspects of the odorant response, e. g. the role of PDE 1C in response termination \cite{CygnarZhao_PDE_NatNEurosc2009,BoccaccioEtal_PDE2006}. It can be applied to clarify the effect of cAMP clearance via diffusion into the cell body versus PDE hydrolysis by PDE 1C. In addition, our model can be adapted to study other biological systems with constrained spaces, e.g. outer dendrite and sensilla in insect olfaction, microvilli of insect photoreceptors, taste cells, or synapses and synaptic buttons \cite{BookFainSensoryTransduction}. Each of these systems has to balance its functional need for ionic currents with the advantages or limitations of having to operate with a small intracellular volume. In olfactory cilia, this is achieved by combining a cationic current with a secondary anionic current to ensure response amplification that preserves response stability and reliability. 

\section{Materials and Methods}

\subsection{Single cell recordings of odorant-induced responses}
Mice were handled and euthanized in accordance with methods approved by the Monell Chemical Senses Center Institutional Use and Care Committee. Mice were euthanized with CO$_2$ followed by cervical dislocation. The suction-pipette technique \cite{LoweGold_1991,PonisseryEtal_2012} was used to record odorant-induced responses from isolated ORNs. In short, the cell body of an isolated ORN is sucked into the tip of a recording pipette. The cilia remain outside of the pipette and therefore accessible for solution changes and odorant stimulation. No access is gained to the intracellular environment, thus intracellular ion concentration remain unperturbed. Also, the intracellular voltage is free to vary, as the ORN is not voltage-clamped. The current recorded in this configuration is the current that enters via the ciliary transduction channels and exits via the cell body. Currents were recorded with a Warner PC-501A patch clamp amplifier (Warner Instrument), filtered DC - 50 Hz and digitized at 10 kHz using a Power1401 MK2 acquisition board and Signal software (Cambridge Electronic Design). Fast solution exchanges were achieved by moving the ORN in the tip of the recording pipette across the interface to two parallel stream of solution using the Perfusion Fast-Step system (Warner Instruments). Ringer solution contained (in mM): 140 NaCl, 5 KCl, 1 $MgCl_2$, 2 $CaCl_2$, 0.01 EDTA, 10 HEPES, and10 glucose. The pH was adjusted to 7.5 with NaOH. ORNs were stimulated with an odorant mix of cineole and acetophenone. 

\subsection{Computational model}
We constructed a spatially-resolved mathematical model to analyze odorant responses in vertebrate ORNs. Our model accounts for the biochemical transduction pathway (Fig.~\ref{FigIntroduction}B) and the spatio-temporal ion dynamics in the cilium during an odorant response. A schematic of the electrical part of our ciliary model is shown in Fig.~\ref{FigSimAveraged}A, for details we refer the interested reader to the SI.

\section{Acknowledgements}
We thank G. Fain, C-Y. Su, G. Lowe, M. Zapotocky and P. Lucas for helpful comments on the manuscript. This work was supported by a National Institutes of Health grant to J. Reisert (R01DC016647). J. Reingruber would like to thank the Isaac Newton Institute for Mathematical Sciences, Cambridge, for support and hospitality during the programme {\it Stochastic dynamical systems in biology: numerical methods and applications} where work on this paper was undertaken. This work was supported by EPSRC grant no EP/K032208/1.

\cleardoublepage
\section*{Supplementary Information}
\begin{appendix}

\captionsetup[figure]{labelfont={bf},name={Fig.} ,labelsep=period}
\renewcommand{\thefigure}{S\arabic{figure}}
\renewcommand{\thetable}{S\arabic{table}}
\captionsetup[table]{labelfont={bf},name={Table} ,labelsep=period}
\setcounter{figure}{0} 

%%%%%%%%%%%%%%%%%%%%
\section{Derivation of the spatially resolved model}
%%%%%%%%%%%%%%%%%%%%

The initial steps in the signal transduction pathway occur in long and slender cilia, for reviews see \cite{Reisert_Review2005,Kleene_Review2008,SchildRestrepo_Review_1998}. The activation of an odorant receptor in the ciliary membrane initiates a G protein-coupled cascade that leads to the opening of CNG channels. The resulting $Ca^{2+}$ influx activates $Ca^{2+}$-activated $Cl^-$ channels (Ano2) that further amplify the response. Ciliary cAMP is degraded by phosphodiesterases and $Ca^{2+}$ is removed by $Na^+/K^+/Ca^{2+}$ exchangers (NCKX4) \cite{StephanReisertetal_NCKX4_NatNeurosc2011}. This leads to closure of CNG and Ano2 channels and response termination.

\subsection{Electrodiffusion model for the ion dynamics in a cilium}

To study the ion dynamics in a cilium, we use an electrodiffusion model similarly to the models in  \cite{QianSejnowski_DendriteModel_1989,Morietal_EpahapticConductionElectrodiffusion_PNAS2008,HalnesEtal_PCompBiol2013}. We model a cilium as a rotationally symmetric cylinder of length $L_{ci}$ and radius $R_{ci}$ with membrane surface $S_{ci}=2\pi R_{ci} L_{ci}$, cross section $A_{ci}=\pi R_{ci}^2$ and cytoplasmic volume $V_{ci}=A_{ci} L_{ci}$. We neglect the ciliary space occupied by axoneme since it would not qualitatively change our results. We assume that a cilium is rotationally symmetric and radially homogeneous and we introduce the dimensionless position $z=x/L_{ci}$. Because $R_{ci}\ll L_{ci}$, we assume that a cilium is radially homogeneous. Hence, ciliary quantities depend only on the time and the longitudinal position $0\le x \le L_{ci}$. Instead of $x$ we use the dimensionless position $z=x/L_{ci}$. The cilium tip is at $z=0$ and the cilium base at $z=1$ is the connection to the dendritic knob. We further introduce the dimensionless voltage $\phi=\frac{U}{U_T}$, where $U_T=\frac{RT} {{\cal F}}\approx 25 mV$ at $T=293K$ (${\cal F}$ is the Faraday constant and $R$ the ideal gas constant). We consider the dynamics of $K^+$, $Na^+$, $Ca^{2+}$ and $Cl^{-}$. We assume that negatively charged organic anions are immobile. As a consequence, no currents are associated with them and they do not contribute to the dynamic. However, it is straightforward to include their dynamics in a more detailed version of our model.  

Ions of valence $z_s$ diffuse with diffusion constant $D_s$, where $s={ca,cl,na,k}$ labels the ion species. The ciliary concentrations $c^s_{ci}(z,t)$ change due to membrane fluxes $J^{s}_{ci,mu}(z,t)$ and longitudinal fluxes $J^{s}_{ci}(z,t)$. Fluxes depend on channel and exchanger properties, and concentration and voltage gradients. We introduce the modified fluxes 
\begin{eqnarray}\label{redefinedFluxes}
\begin{array}{rcl}
\displaystyle j^{s}_{ci,mu}&=& \displaystyle \frac{dS_{ci}}{dV_{ci}} J^{s}_{ci,mu}= \frac{2}{R_{ci}}J^{s}_{ci,mu}\\ \\
\displaystyle j^{s}_{ci} &=& \displaystyle \frac{J^{s}_{ci}}{L_{ci}}   = -\nu^s_{ci} \left( \partial_z c^s_{ci}+ z_s c^s_{ci} \partial_z \phi_{ci}\right)\,,
\end{array}
\end{eqnarray}
where $\nu^s_{ci} =\frac{D_s}{L_{ci}^2}$ are diffusional rate constants. Note that $j^{s}_{ci,mu}$ and $j^{s}_{ci}$ have unit $\frac{mM}{s}$. The flux through a surface element $dS_{ci}$ is $J^s_{ci,mu}dS_{ci} =  j^s_{ci,mu}dV_{ci} = V_{ci} j^s_{ci,mu}dz$, and the longitudinal flux is $A_{ci} J^{s}_{ci} = V_{ci} j^{s}_{ci}$. The modified fluxes have intuitive interpretations:  during one second, a uniform flux would change the ciliary concentration by $j$. The currents associated with the fluxes are
\begin{eqnarray}\label{defCurrents}
\begin{array}{rcl}
\displaystyle I^s_{ci,mu} &=&   z_s {\cal F} V_{ci} j^{s}_{ci,mu} \\ \\
\displaystyle I^s_{ci}  &=& \displaystyle z_s   {\cal F}  V_{ci} j^{s}_{ci} \,.
\end{array}
\end{eqnarray}
With $V_{ci}\sim 0.5 \mu m^3$ we find ${\cal F} V_{ci}\sim 0.05 \frac{s \, pA}{mM}$. Thus, a  current of 1pA corresponds to a flux of 1/0.05=20 mM/s. Hence, in one millisecond a current of 100pA can, in principle,  change the ciliary concentration by 2 mM. This illustrates that currents can quickly alter the ciliary concentrations. 

We apply the usual convention that inward fluxes are negative and outward fluxes are positive. Thus, and inward flux of cations or an outward flux of anions generate a negative current. The conservation equations (Nernst-Planck equations) for the ion concentrations are \cite{BookJackNobleTsien}
\begin{eqnarray}\label{eqForConcCi}
\partial_t c^s_{ci}=  - \partial_zj^{s}_{ci} - j^{s}_{ci,mu}\,.
\end{eqnarray}

We simplify and ignore $Ca^{2+}$ buffering and we only use the rapid buffer approximation where $Ca^{2+}$ diffusion is slowed down due to rapid bindings and unbindings. However, for the sake of completeness, we present here also equations for the $Ca^{2+}$ dynamics with exchange with fast and slow $Ca^{2+}$ buffers that are immobile. The dynamics of the slow calcium buffer $c^{cab}_{ci}(z,t)$ is 
\begin{eqnarray}
\partial_t c^{cab}_{ci}= \beta_{cab} \left(  ( c^{cab}_{ci,tot} -c^{cab}_{ci}) \frac{c^{ca}_{ci}}{K_{cab}} -  c^{cab}_{ci} \right) 
\end{eqnarray}
where $c^{cab}_{ci,tot}$ is the total slow buffer concentration. The amount of $Ca^{2+}$ that is stored in fast buffers is $B_{ca} c^{ca}_{ci}$, where $B_{ca}$ is the buffering constant. The overall amount of ciliary $Ca^{2+}$ changes due to flux of free $Ca^{2+}$ such that $\partial_t\left( (1+B_{ca}) c^{ca}_{ci} + c^{cab}_{ci}\right)  = -\partial_z j^{ca}_{ci}- j^{ca}_{ci,mu}$. Thus, the free  $Ca^{2+}$ dynamics is
\begin{eqnarray}\label{eqforCaWithBuf}
\partial_t c^{ca}_{ci} = \frac{1}{1+B_{ca}} \left( -\partial_z j^{ca}_{ci} - j^{ca}_{ci,mu} -\partial_t c^{cab}_{ci}\right) \,.
\end{eqnarray}

Due to the large mucosal volume, we neglect concentration and potential changes in the mucus during an odorant response. However, we do explore the impact of different values for the mucus concentrations. We use the mucus potential as reference and we set  $\phi_{mu}=0$ such that $\phi_{ci} - \phi_{mu} =\phi_{ci}$. 

A challenge is to compute the spatio-temporal dynamics of the ciliary potential $\phi_{ci}(z,t)$. The most accurate way would be to solve the Poisson equation with appropriate boundary conditions, leading to the Poisson-Nernst-Planck (PNP) coupled system of equations. However, such a complex system of partial differential equations, with additional feedback from the cell body dynamics, would be very demanding for numerical simulations. In particular, we are interested in simulations that lasts for seconds and not only milliseconds or steady state results \cite{GardnerEtal_2015,Podsetal_PNP_BJ2013,Sejnowskietal_NodeRanvier_BJ2008}. To reduce complexity, we use the charge-capacitor approximation to compute the potential as a function of the electric currents and the membrane capacitance
\cite{QianSejnowski_DendriteModel_1989,Mori_ElectrodiffusionOsmosis_PhysicaD2015,HalnesEtal_PCompBiol2013}
\begin{eqnarray}\label{eqForPotentialCiliaFlux}
\displaystyle  \partial_t  \phi_{ci}  = \frac{1}{C_{ci} U_T} \sum_s  \left(- \partial_z  I^{s}_{ci} -  I^{s}_{ci,mu} \right)\,.
\end{eqnarray}
$C_{ci} = S_{ci} C_m$ is the cilium capacitance ($C_m=1{\mu F}/{cm^2}= 10^{-5} {nF}/{\mu m^2}$ is the membrane capacity). Eq.~\ref{eqForPotentialCiliaFlux} is formally identical to a cable equation \cite{BookKeenerSneyd}. 
%With Eq.~\ref{eqForConcCi} we can rewrite  Eq.~\ref{eqForPotentialCiliaFlux} as $\frac{\partial}{\partial t}\phi_{ci}= \frac{ V_{ci} {\cal F} } { C_{ci} U_{T}} \frac{\partial}{\partial t}  \sum_s z_s c^s_{ci}$, and after integration this becomes  $\phi_{ci}=  \frac{ V_{ci} {\cal F} } { C_{ci} U_{T}}  \sum_s z_s c^s_{ci} +   \psi_{ci}(z)$. The constant organic anion concentration contributes to $\psi_{ci} (z)$.

%
\subsection{Well-stirred model for the cell body}
Due to the large size of the cell body, we approximate the cell body concentrations $c_{cb}^s$ as being constant. It is straightforward to relax this assumption in a more sophisticated future model. The cell body potential $\phi_{cb}(t)$ changes due to a leak current, which defines the resting potential, and the currents from $N_{ci}$ identical cilia, 
\begin{eqnarray}\label{eqCellBodyPotLeak}
\frac{\partial }{\partial t} \phi_{cb}=  \frac{N_{ci}}{C_{cb} U_T} \sum_{s}  I^{s}_{ci,cb} - \frac{g_{cb}^{leak}}{C_{cb}} \left(\phi_{cb}- \phi^{leak}_{cb}\right)  \,.
\end{eqnarray}
$C_{cb}$ is the cell body capacitance, $g_{cb}^{leak}$ is the leak conductance, $\phi^{leak}_{cb}$ is the leak reversal potential and $I^{s}_{ci,cb}(t)= I^{s}_{ci}(1,t) = z_s {\cal F} V_{ci}  j^s_{ci}(1,t)$ is the current from a cilium carried by ion species $s$. 
%The steady state potential is 
%\begin{eqnarray}
%	\phi_{cb}= \phi^{leak}_{cb} +  \frac{N_{ci}}{U_T g_{cb}^{leak} }\sum_{s}  I^{s}_{ci,cb}\,.
%\end{eqnarray}

%%%%%%%%%%%%
\subsection{Ion fluxes between cilium and mucus }
%%%%%%%%%%%%
The fluxes between cilium and mucus depend on CNG and Ano2 channels and electrogenic NCKX4 exchangers
\begin{eqnarray}
j^{s}_{ci,mu}= j^{s,cng}_{ci,mu}+ j^{s,ano}_{ci,mu} + \zeta^s_{nckx} j^{nckx}_{ci,mu}\,.
\end{eqnarray}
The numbers $\zeta^s_{nckx}$ define the exchange stoichiometry. 
\subsubsection{Fluxes via CNG and Ano2 channels}
For the channel fluxes between cilia and mucus we use the Goldman-Hodgkin-Katz (GHK) flux equation \cite{BookKeenerSneyd}. With $\phi_{mu}=0$ such that $\phi_{ci,mu}=\phi_{ci}-\phi_{mu}=\phi_{ci}$ and the approximation 
\begin{eqnarray}
\frac{x}{e^{x/2}-e^{-x/2}}\approx e^{-\frac{x^2}{24}}
\end{eqnarray}
valid for $|x| \lesssim 5$ we have 
\begin{eqnarray} 
j^{s,ch}_{ci,mu}  &=&   p_{ch}  \nu^s_{ch}e^{-\frac{(z_s \phi_{ci})^2}{24}}   \left( c^s_{ci} e^{z_s \phi_{ci}/2} -c^s_{mu} e^{-z_s \phi_{ci}/2}\right) \label{GHKcurrent}\\
&=& p_{ch}  \nu^s_{ch}e^{-\frac{(z_s \phi_{ci})^2}{24}}  c^s_{ci} e^{z_s \phi_{ci}/2} \left(1 - \frac{c^s_{mu}}{c^s_{ci} } e^{-z_s \phi_{ci}}\right)  \label{GHKcurrentNernst1} \\
&=&   -p_{ch}  \nu^s_{ch}e^{-\frac{(z_s \phi_{ci})^2}{24}}   c^s_{mu} e^{-z_s \phi_{ci}/2} \left( 1- \frac{c^s_{ci}}{c^s_{mu} } e^{z_s \phi_{ci}} \right) \,.  \label{GHKcurrentNernst2}
\end{eqnarray}
$p_{ch}(z,t)$ is the channel open probability and the rate $\nu^s_{ch}\sim \frac{\rho^s_{ch}\gamma^s_{ch}}{R_{ci}}$ is proportional to the channel density $\rho_{ch}$ and the single channel conductivity $\gamma^s_{ch}$ and $R_{ci}^{-1}$ (due to Eq.~\ref{redefinedFluxes}). 

Eq.~\ref{GHKcurrentNernst1} shows that $Cl^-$ efflux is proportional to the ciliary $Cl^-$ concentration, and altering the mucosal $Cl^-$ concentration affects this efflux only only by little via the term $1 - c^{cl}_{mu}/c^{cl}_{ci} e^{\phi_{ci}}$. In contrast, Eq.~\ref{GHKcurrentNernst2}  shows that $Na^+$ influx is proportional to the mucosal $Na^+$ concentration and is therefore strongly affected by mucosal concentration changes (we have $1 - c^{na}_{ci}/c^{na}_{mu} e^{\phi_{ci}}\approx 1$) . 

The open probability for Ano2 channels $p_{ano}(z,t)$ depends on the $Ca^{2+}$ concentration \cite{KleeneeGestland_ChlorideConductanceFrog_JNsc1991,Reisertetal_ClCurrent_JGenPhys2003},
\begin{eqnarray}\label{eqAno2Prob}
p_{ano}= \frac{(c^{ca}_{ci})^{h_{ano}}} {(c^{ca}_{ci})^{h_{ano}} + {K_{ano}}^{h_{ano}}} \,,
\end{eqnarray}
with $K_{ano}^{ca}\approx 1.8\mu M$ and $h_{ano}\approx 2.3$ \cite{Reisertetal_ClCurrent_JGenPhys2003}. The CNG open probability $p_{cng}(z,t)$ depends on the cAMP concentration $c_{camp}(z,t)$ \cite{NakamuraGold_CNG_Nature1987,FiresteinZufallShepherd_CNG_JNsc1991}. Furthermore, CNG channels become desensitized due to negative $Ca^{2+}$ feedback via calmodulin (CaM) \cite{ChenYau_CNG_Nature1994,BradleyEtal_CNG_Science2001}. In a first approximation, we incorporate this negative feedback using a $Ca^{2+}$-dependent Hill concentration $K_{cng}(z,t)$,  
\begin{eqnarray}
\displaystyle p_{cng} = \displaystyle  \frac{(c^{camp}_{ci})^{h_{cng}}}{(c^{camp}_{ci})^{h_{cng}} + {K_{cng}}^{h_{cng}}} 
\end{eqnarray}
with   
\begin{eqnarray}
\displaystyle K_{cng}= \displaystyle K_{cng}^{min} \, \left( 1 + f_{cng}^{ca} \frac{c^{ca}_{ci}}{c^{ca}_{ci} + K_{cng}^{ca}}\right)\,.
\end{eqnarray}

\subsubsection{Fluxes via NCKX4 exchangers }
During one cycle, 4 Na$^+$  are exchanged for one K$^+$ and one Ca$^{2+}$ ($\zeta^{na}_{nckx} = -4$, $\zeta^{ca}_{nckx}= \zeta^{k}_{nckx} = 1$). Hence, NCKX4 is an electrogenic exchanger and the net current is
\begin{eqnarray}
I^{nckx}_{ci,mu}=  {\cal F} V_{ci} \sum_s z_s  \zeta^s_{nckx}  j^{nckx}_{ci,mu}   =r  {\cal F} V_{ci} j^{nckx}_{ci,mu}\,,
\end{eqnarray}
where $r =\sum_s z_s  \zeta^s_{nckx} =-1$. The flux $j^{nckx}_{ci,mu}$ corresponds to either Ca$^{2+}$ or K$^+$. To obtain a formula for  $j^{nckx}_{ci,mu}(z,t)$ we generalize the expression for the NCX exchanger \cite{Lindemann_BJ2001,DiFrancescoNoble_1985,Mullins_NaCaTransport_1977}, 
\begin{eqnarray}
j^{nckx}_{ci,mu}&=& \nu_{nckx} \frac{ c^{ca}_{ci} (c^{na}_{mu})^4c^{k}_{ci} e^{r\phi_{ci}/2} - c^{ca}_{mu}(c^{na}_{ci})^4 c^{k}_{mu} e^{-r\phi_{ci}/2 } } {(c^{ca}_{ci}+K_{nckx}) (c^{na}_{mu})^4  c^{k}_{ci}  + (c^{ca}_{mu}+K_{nckx}) (c^{na}_{ci})^4 c^{k}_{mu}} \label{exchangerFlux} \\
&=& \nu_{nckx}  c^{ca}_{ci} (c^{na}_{mu})^4c^{k}_{ci} e^{r\phi_{ci}/2} 
\frac{1 -e^{- r\phi_{ci}  + \sum_s z_s \zeta^s_{nckx} \phi^{s}_{ci,mu} } }
{(c^{ca}_{ci}+K_{nckx}) (c^{na}_{mu})^4  c^{k}_{ci}  + (c^{ca}_{mu}+K_{nckx}) (c^{na}_{ci})^4 c^{k}_{mu}} \nonumber 
%&=&   \nu_{nckx} e^{r\phi_{ci}/2}  \frac{ c^{ca}_{ci} - c^{ca}_{mu}\left(\frac{c^{na}_{ci}}{c^{na}_{mu}}\right)^4 \frac{c^{k}_{mu}}{c^{k}_{ci}} e^{-r\phi_{ci} } } { K_{nckx} + c^{ca}_{ci} +  (c^{ca}_{mu}+K_{nckx})\left(\frac{c^{na}_{ci}}{c^{na}_{mu}}\right)^4 \frac{c^{k}_{mu}}{c^{k}_{ci}}  }  \,
\end{eqnarray}
with the Nernst potentials 
\begin{eqnarray}\label{defNernstPotential}
\phi^s_{ci,mu}  = \frac{1}{z_s} \ln \left(\frac{c^s_{mu}}{c^s_{ci}} \right)\,.
\end{eqnarray}
$j^{nckx}_{ci,mu}$ depends symmetrically on ciliary and mucosal parameters. The flux vanishes when the free energy change  is zero, $\Delta E \sim \sum_s z_s \xi^s_{nckx} ( \phi^{s}_{ci,mu} - \phi_{ci}) =0$. The ratio between ciliary and mucus $Ca^{2+}$  concentration at this point is 
\begin{eqnarray}\label{exchangerCaReduction}
\frac{c^{ca}_{ci}}{ c^{ca}_{mu}} =\left ( \frac{c^{na}_{ci}}{c^{na}_{mu}} \right )^4 \frac{c^{k}_{mu}}{c^{k}_{ci}} \,e^{-r\phi_{ci}} \,.
\end{eqnarray}
At rest, with $\phi_{ci}=-2.4$ (corresponding to -65mV), $c^{na}_{mu}=140mM$, $c^{na}_{ci}=4mM$,  $c^{k}_{mu}=5mM$, $c^{k}_{ci}=140mM$ we find $\frac{c^{ca}_{ci}}{ c^{ca}_{mu}}\sim 10^{-9}$.  Thus, with $c^{ca}_{mu}= 2mM$ and no $Ca^{2+}$ influx into cilia, the exchangers are able to reduce the ciliary $Ca^{2+}$ concentration much below nanomolar level. For $c^{ca}_{ci}  \gg c^{ca}_{mu}\left(\frac{c^{na}_{ci}}{c^{na}_{mu}}\right)^4 \frac{c^{k}_{mu}}{c^{k}_{ci}}$ Eq.~\ref{exchangerFlux} can be approximated by the Michaelis-Menton form 
\begin{eqnarray}
j^{nckx}_{ci,mu} \approx  \nu_{nckx} e^{r\phi_{ci}/2} \frac{c^{ca}_{ci} } {K_{nckx}  +  c^{ca}_{ci} } \,.
\end{eqnarray}

\subsection{Signal transduction and cAMP dynamics}
Since the focus of this paper is on the ion dynamics, we use a basic transduction model comprising odorant receptor, G protein, adenylate cyclase (AC), cAMP  and CaMK \cite{DePaloetal_BJ2012,Reidletal_CaOsc_BJ2006,Doughertyetal_PNAS2005}.  After the activation of odorant receptors (or), a G protein (g) coupled transduction cascade leads to the activation of adenylate cyclase (ac) that synthesizes cAMP. We neglect diffusion of receptors, G-protein and AC in the membrane. Because experimental evidences suggests that the lifetime of an activated odorant receptor is less than a millisecond \cite{GhatpandeReisert_JPhys2011,BhandawatReisertYau_Science2005}, we use the steady state expression to model the fraction (with respect to the total concentration) of activated receptors $or^*$ as a function of a slowly varying odorant stimulus $od(z,t)$, $or^*= \frac{od^{h_{od}}}{od^{h_{od}}+K_{od}^{h_{od}}}$. The equations for the fractions of activated G protein and AC are 
\begin{eqnarray}
\begin{array}{rcl}
\displaystyle \partial_t g^* &=& \displaystyle \beta_{g} \left( (1-g^*) \frac{ or^*}{K_{or}}-g^*\right)   \\ \\
\displaystyle  \partial_t ac^* &=&  \displaystyle  \beta_{ac} \left( (1-ac^*) \frac{ g^*}{K_{g}} -  ac^* \right) 
\end{array}
\end{eqnarray}
where  $0\le g^*(z,t)\le 1$ and $0\le ac^*(z,t)\le 1$.  The rate constants $\beta_{g}$ and $ \beta_{ac} $ determine the activation dynamics. Steady state fractions are $g^*= \frac{or^*}{or^* +K_{or}}$ and $ac^* =  \frac{g^*}{g^* + K_g}$.  

The synthesis of cAMP is stimulated by $ac^*$ and is inhibited by $Ca^{2+}$ via  CaMK feedback. In first approximation, we model the CaMK feedback directly as a function of  $Ca^{2+}$ using and effective Hill equation. Thus, the cAMP concentration $c^{camp}_{ci}(z,t)$ satisfies the reaction-diffusion equation 
\begin{eqnarray}\label{eqcAMP}
\partial_t c^{camp}_{ci} =\displaystyle - \partial_z j^{camp}_{ci}  + ac^* \frac{ \alpha^{max}_{camp} }{ 1+f_{camk}}   -  \beta_{camp} c^{camp}_{ci} 
\end{eqnarray}
with  $ j^{camp}_{ci}=-\nu^{camp}_{ci} \partial_z c^{camp}_{ci}$ and $\nu^{camp}_{ci}=\frac{D_{camp}}{L_{ci}^2}$. The dimensionless parameter $f_{camk}$  accounts for $Ca^{2+}$ dependent inhibition of cAMP synthesis due to CaMK feedback,  
\begin{eqnarray}
\begin{array}{rcl}
\displaystyle \frac{d}{dt}f_{camk}=  \displaystyle  \beta_{camk}\left( \frac{f_{camk}^{max}}{1+\left(\frac{K_{camk}}{c^{ca}_{ci}}\right)^{h_{camk}}} - f_{camk}\right) \,.
\end{array}
\end{eqnarray}
%The flux boundary condition at $z=1$  is 
%\begin{eqnarray} \label{eqcAMPFluxCiCb}
%	j^{camp}_{ci,cb} = \alpha_{ci,cb} \nu^{camp}_{ci}  (c^{camp}_{ci}(1,t) - c^{camp}_{cb} ) \,.
%\end{eqnarray}

%%%%%%%%%%%%%%%%%%%%%%%%%%%%%%%%%%%%%%%%%%%%%%%%%%%%%%%%%%%%%
\subsection{Ciliary boundary conditions }
%%%%%%%%%%%%%%%%%%%%%%%%%%%%%%%%%%%%%%%%%%%%%%%%%%%%%%%%%%%%%%
The tip of a cilium at $z=0$ is a sealed end with reflecting boundary conditions $j^{s}_{ci}(0,t) = 0$. At $z=1$ the cilia are connected via the dendritic knob and the dendrite to the cell body. To reduce complexity, we refrain from implementing a spatial model for electrodiffusion in knob and dendrite. Instead, we use flux boundary conditions $j^{s}_{ci}(1,t)=j^{s}_{ci,cb}(t)$ where $j^{s}_{ci,cb}$ depends on concentration and voltage differences between cilia and cell body and on the ciliary and dendritic geometry. We do not use Dirichlet conditions $c^s_{ci}(1,t)=c^s_{cb}=constant$ because flux boundary conditions provide several advantages: First, flux conditions are flexible and allow to account for geometry diffusional constraints. Second, Dirichlet conditions can be retrieved from flux boundary conditions by changing parameters (see Fig.~\ref{FigAlphaCiCb}). Finally, with flux boundary conditions it is straightforward to map our spatial model into a well-stirred model that we use for parameter estimations (see next paragraph). In contrast, it is unclear how to consistently map a spatial model with constant Dirichlet boundary conditions into a well-stirred model with time dependent concentrations.  

Because the diameters of knob and dendrite are not much different \cite{MorrisonConstanzo_Review1992}, we integrate the knob into the dendrite. We further assume that the dendrite acts like a radially well-stirred and homogeneous cylinder with no current leakage and voltage that changes linearly between cilia and the cell body. As a first approximation, we therefore use for $j^{s}_{ci,cb}$ an expression that is similar to the GHK formula for the flux through a channel \cite{BookKeenerSneyd}, 
\begin{eqnarray} \label{GHKfluxCiCb}
j^{s}_{ci,cb} = \alpha_{ci,cb} \nu^s_{ci}  e^{-\frac{(z_s \phi_{ci,cb})^2}{24}} \left(c^s_{ci}(1,t) e^{z_s\phi_{ci,cb}/2} - c^s_{cb} e^{-z_s\phi_{ci,cb}/2}\right),
\end{eqnarray}
with $\phi_{ci,cb}(t) =\phi_{ci}(1,t)  - \phi_{cb}(t)$. The corresponding flux boundary condition for the ciliary potential $\phi_{ci}$ is 
\begin{eqnarray}
j^\phi_{ci,cb}  = \frac{ 1}  { C_{ci} U_{T}}   \sum_s  z_s I^{s}_{ci,cb}  = \frac{ V_{ci} {\cal F} } { C_{ci} U_{T}}   \sum_s  z_s  j^{s}_{ci,cb}  .
\end{eqnarray}
The dimensionless parameter $\alpha_{ci,cb}$ accounts for ciliary and dendritic properties. For example, for small $\alpha_{ci,cb}$ the flux between cilium and cell body is strongly restricted, whereas for large $\alpha_{ci,cb}$ we recover Dirichlet boundary conditions (see Fig.~\ref{FigAlphaCiCb}). To obtain a first estimation of $\alpha_{ci,cb}$, we assume that the overall dendritic flux $J^{s}_{dend}$ can also be approximated as 
\begin{eqnarray} 
J^{s}_{dend} = A_{dend} \frac{D_s}{L_{dend}} e^{-\frac{(z_s \phi_{ci,cb})^2}{24}}  \left(c^s_{ci}(1,t) e^{z_s\phi_{ci,cb}/2} - c^s_{cb} e^{-z_s\phi_{ci,cb}/2}  \right)  \,.
\end{eqnarray}
where $L_{dend}$ and $A_{dend}=\pi R_{dend}^2$ are dendritic length and cross-section. With current conservation $J^{s}_{dend}=N_{ci} j^{s}_{ci,cb}V_{ci}$ and Eq.~\ref{GHKfluxCiCb} we find  
\begin{eqnarray}\label{eqForAlphaCiCb}
\alpha_{ci,cb} =  \frac{1}{N_{ci}} \left ( \frac{R_{dend} }{R_{ci}}\right )^2  \frac{ L_{ci}}{ L_{dend}} \,.
\end{eqnarray}
Eq.~\ref{eqForAlphaCiCb} has reasonable properties: ciliary efflux is facilitated for large $A_{dend}$ and it is obstructed for large $L_{dend}$; if many cilia converge into a dendrite, the efflux from a single cilium becomes restricted due to limitations of the flux through the dendrite. With $N_{ci}\sim 15$, $R_{dend}/R_{ci}\sim 10$  and $L_{ci}/L_{dend}\sim 1$ \cite{MorrisonConstanzo_Review1992} we find $\alpha_{ci,cb} \sim 7$. We used this value for the simulations presented in the main text. However, a different value for $\alpha_{ci,cb}$ would not alter our conclusions from the main text, see Fig.~\ref{FigAlphaCiCb}.

%%%%%%%%%%%%%%%%%%%%%%%%%%%%%%%%%%%%%%%%%%%%%%%%%%%%%%%%%%%%%
\section{Derivation of the well-stirred model }
%%%%%%%%%%%%%%%%%%%%%%%%%%%%%%%%%%%%%%%%%%%%%%%%%%%%%%%%%%%%%
\label{wellStirredSection}
%%%%%%%%%%%%%%%%%%%%%%%%%%%%%%%%%%%%%%%%%%%%%%%%%%%%%%%%%%%%%%
Starting from the spatial model we derived well-stirred equations that we used to fit experimental data and estimate parameters. The well-stirred equations are obtained by replacing $c^s_{ci}(z,t)$ with $c^s_{ci}(t)$ and $\partial_z j^{s}_{ci}(z,t)$ with $j^{s}_{ci,cb}(t)$. Thus, the well-stirred equations for concentrations and potential read  
\begin{eqnarray}\label{eqWellStirred}
\begin{array}{rcl}
\displaystyle  \frac{d}{dt}c^s_{ci}  &=&  \displaystyle - j^{s}_{ci,cb} - j^{s}_{ci,mu} \\ \\
\displaystyle \frac{d}{dt}c^{ca}_{ci} &=& \displaystyle \frac{1}{1+B_{ca}} \left( - j^{ca}_{ci,cb}  - j^{ca}_{ci,mu} -\frac{d}{dt} c^{cab}_{ci}\right) \\ \\
\displaystyle  \frac{d}{dt} c^{camp}_{ci} &=&   \displaystyle  ac^* \frac{ \alpha^{max}_{camp} }{ 1+f_{camk}}   -  \beta_{camp} c^{camp}_{ci} -j^{camp}_{ci,cb}  \\ \\
\displaystyle   \frac{d}{dt}  \phi_{ci}  &=& \displaystyle \frac{ 1} { C_{ci} U_{T}}   \sum_s   \left( -  I^{s}_{ci,cb} -  I^{s}_{ci,mu}\right)  \,.
\end{array}
\end{eqnarray}
The rest of the equations remain formally unchanged.

%%%%%%%%%%%%%%%%%%%%%%%%%%%%%%%%%%%%%%%%%%%%%%%%%%%%%%%%
\section{Fitting procedure and derivation of a common set of parameters}
%%%%%%%%%%%%%%%%%%%%%%%%%%%%%%%%%%%%%%%%%%%%%%%%%%%%%%%%
\begin{figure}[htb]
	\begin{center}	
		\includegraphics[width=0.5\linewidth]{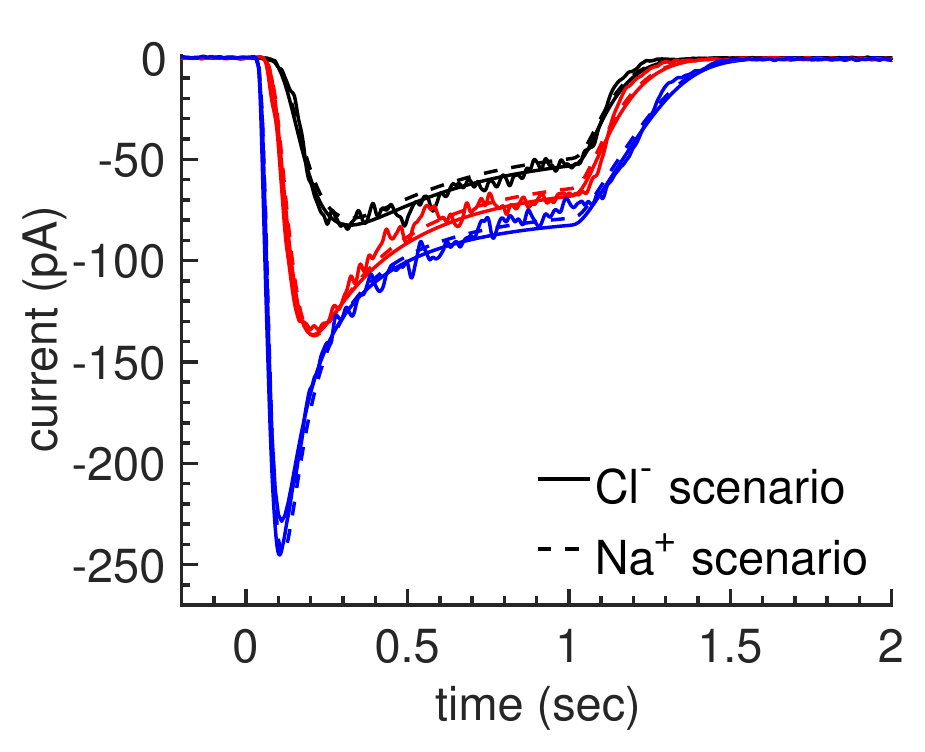}
		\caption{{\bf Fitting with the well-stirred model.  } Odorant-induced responses (noisy traces) from a mouse olfactory receptor neuron to three different one second odorant stimulations with increasing concentrations (10, 30 and 100 $\mu$M) together with fitting results for each scenario.} 
		\label{FigFitting} 
	\end{center}
\end{figure}

In the biological scenario (referred to as  the $Cl^-$ scenario) amplification is due to $Cl^-$ efflux, whereas in the artificial scenario (referred to as the $Na^+$ scenario) it is due to $Na^+$ influx. Because $Na^+$ and $K^+$ currents through CNG channels are small compared to the Ano2 current \cite{Reisertetal_ClCurrent_JGenPhys2003,Billigetal_Ano2_JPhys2011,DibattistaEtal_Review2017}, we simplified and assumed that CNG channels are only permeable to $Ca^{2+}$. To minimize the differences between the two scenarios, in the $Na^+$ scenario we introduce artificial Ano2 channels that have the same $Ca^{2+}$ activation properties as the biological Ano2 channels but are permeable to $Na^+$ instead of $Cl^-$. In this way, both scenarios are absolutely identical except for the Ano2 permeability. To implement a $Ca^{2+}$-activated $Na^+$ conductance for Ano2 channels, we included the rate $\nu^{na}_{ano}$, similarly to $\nu^{cl}_{ano}$ for the $Cl^-$ conductance. Thus, in the $Na^+$ scenario we have ($\nu^{cl}_{ano}>0$, $\nu^{na}_{ano}=0$), whereas in the $Cl^-$ scenario we have ($\nu^{cl}_{ano}=0$, $\nu^{na}_{ano}>0$). Because the driving forces for $Cl^-$ and $Na^+$ are different, also the values for $\nu^{na}_{ano}$ and $\nu^{cl}_{ano}$ have to be different in order to obtain similar peak currents. 

Whenever possible we used published values for our parameters (see Tables \ref{table1} and \ref{table2}). For unknown parameters (e.g. most of the transduction parameters), we estimated values by calibrating the model to suction electrode recordings from a mouse ORN in response to a 1 s stimulation with three different odorant concentrations (Fig.~\ref{FigFitting}, wiggled lines). Our goal was to find a common set of parameters that differs only in the values of $\nu^{na}_{ano}$ and $\nu^{cl}_{ano}$ but equally well fits the experimental data in both scenario. To compute such parameters, we generated two identical datasets by duplicating the experimental data. We then simultaneously fitted both datasets with the constraints $\nu^{na}_{ano}=0$ for dataset 1, and $\nu^{cl}_{ano}=0$ for dataset 2. In this way there was no procedural difference between the two scenarios. Because it is difficult to implement a fitting procedure based on the spatial model, we used the well-stirred model to perform the fitting using the Data to Dynamics (D2D) software \cite{D2D_2013,D2D_2015} based on maximum likelihood estimators. Fig.~\ref{FigFitting} shows that the data was equally well fitted in both scenarios. In a next step, we used the parameters (given in Tables \ref{table1} and \ref{table2}) together with the assumption of homogeneous odorant stimulation along a cilium to perform simulations with our spatial model. We found that the parameters computed with the well-stirred model also provided a reasonably good fit to the data when used with the spatial model (see Fig. 2B in the main text). 

%%%%%%%%%%%%%%%%%%%%%%%%%%%%%%%%%%%%%%%%%%%%%%%%%%%%%%%%
\section{Simulation results}
%%%%%%%%%%%%%%%%%%%%%%%%%%%%%%%%%%%%%%%%%%%%%%%%%%%%%%%%

We now present additional simulation results with strongest odorant stimulation (100 $\mu$M) that complement the results shown from the main text. Parameters are the same as in the main text, except for the ones that are modified as indicated in the figures. 

\begin{figure}[htb]
	\begin{center}	
		\includegraphics[width=0.8\linewidth]{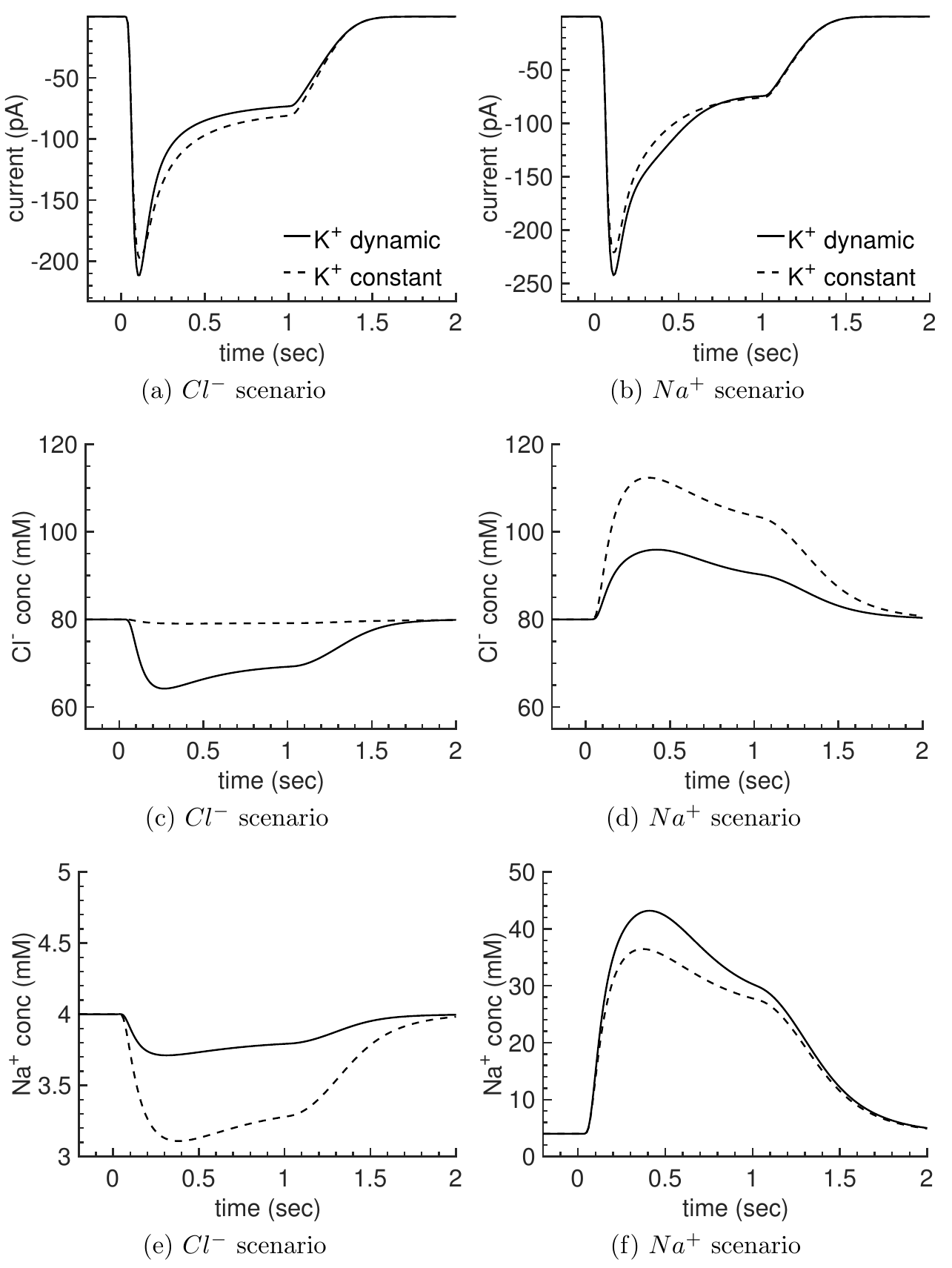}
		\caption{{\bf Dynamic vs constant ciliary $K^+$ concentration.} Comparison of simulation results with dynamic (solid lines) and fixed (dashed lines) $K^+$ concentration performed with the spatial model and strongest odorant stimulation (100 $\mu$M). The panels compare spatially averaged results for current, $Cl^-$ and $Na^+$ concentrations  (see also Fig. 3 in the main text). } 
		\label{FigPotassium} 
	\end{center}
\end{figure}

In Fig.~\ref{FigPotassium} we compare simulation results with dynamic $K^+$ concentration (black lines, see also Fig. 3 in the main text) to results with fixed $K^+$ concentration (dashed lines). The current is not much affected by holding $K^+$ constant (Fig.~\ref{FigPotassium} a-b), but the dynamics of the $Cl^-$ and $Na^+$ concentrations are strongly altered (Fig.~\ref{FigPotassium}c-f).  For example, in the $Cl^-$ scenario with fixed $K^+$, $Cl^-$ efflux does not entail  a strong reduction in the ciliary $Cl^-$ concentration due to electroneutrality requirements. Thus, affecting the $K^+$ dynamics also strongly affects the dynamics of $Cl^-$ and $Na^+$, showing that the ciliary ion dynamics are strongly coupled.   

\begin{figure}[htb]
	\begin{center}	
		\includegraphics[width=0.8\linewidth]{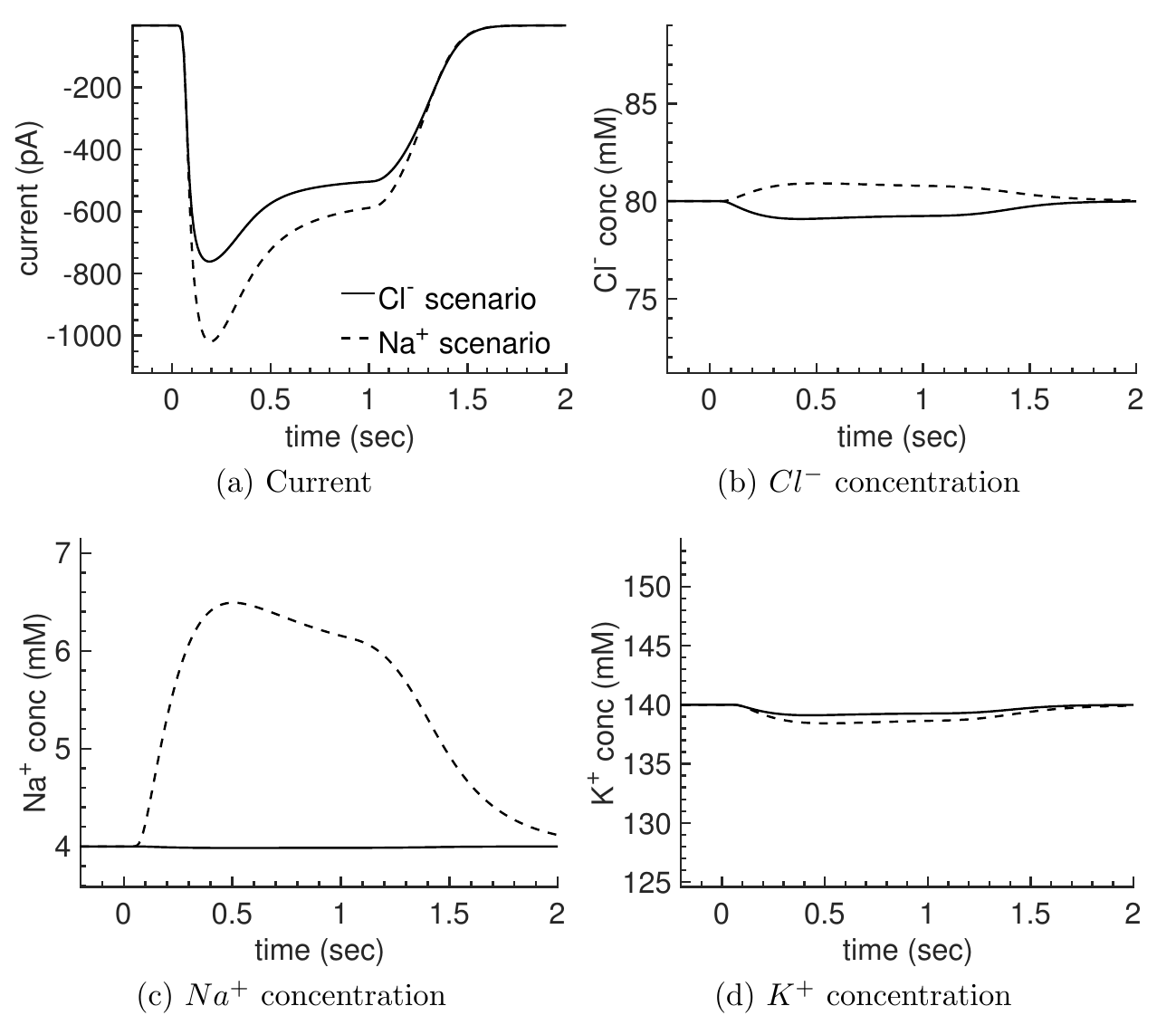}
		\caption{{\bf Impact of a 10-fold increased ciliary radius.} Simulations with 10-fold increased ciliary radius performed with the spatial model and strongest odorant stimulation. The panels compare spatially averaged results for current and ion concentrations between the two scenarios.} 
		\label{FigRadius} 
	\end{center}
\end{figure}

In Fig.~\ref{FigRadius} we compare simulation results between the two scenarios with the ciliary radius that is increased from 75nm to 750nm. Although the current is now strongly augmented due the 10-fold increased membrane surface, concentration changes are small due to the 100-fold larger volume. An effective model assuming constant ion concentrations would now be a valid approximation. Furthermore, with large radius the differences between the two scenarios are small and the advantages for amplification based on a $Cl^-$ current disappear. 

\begin{figure}[htb]
	\begin{center}	
		\includegraphics[width=0.65\linewidth]{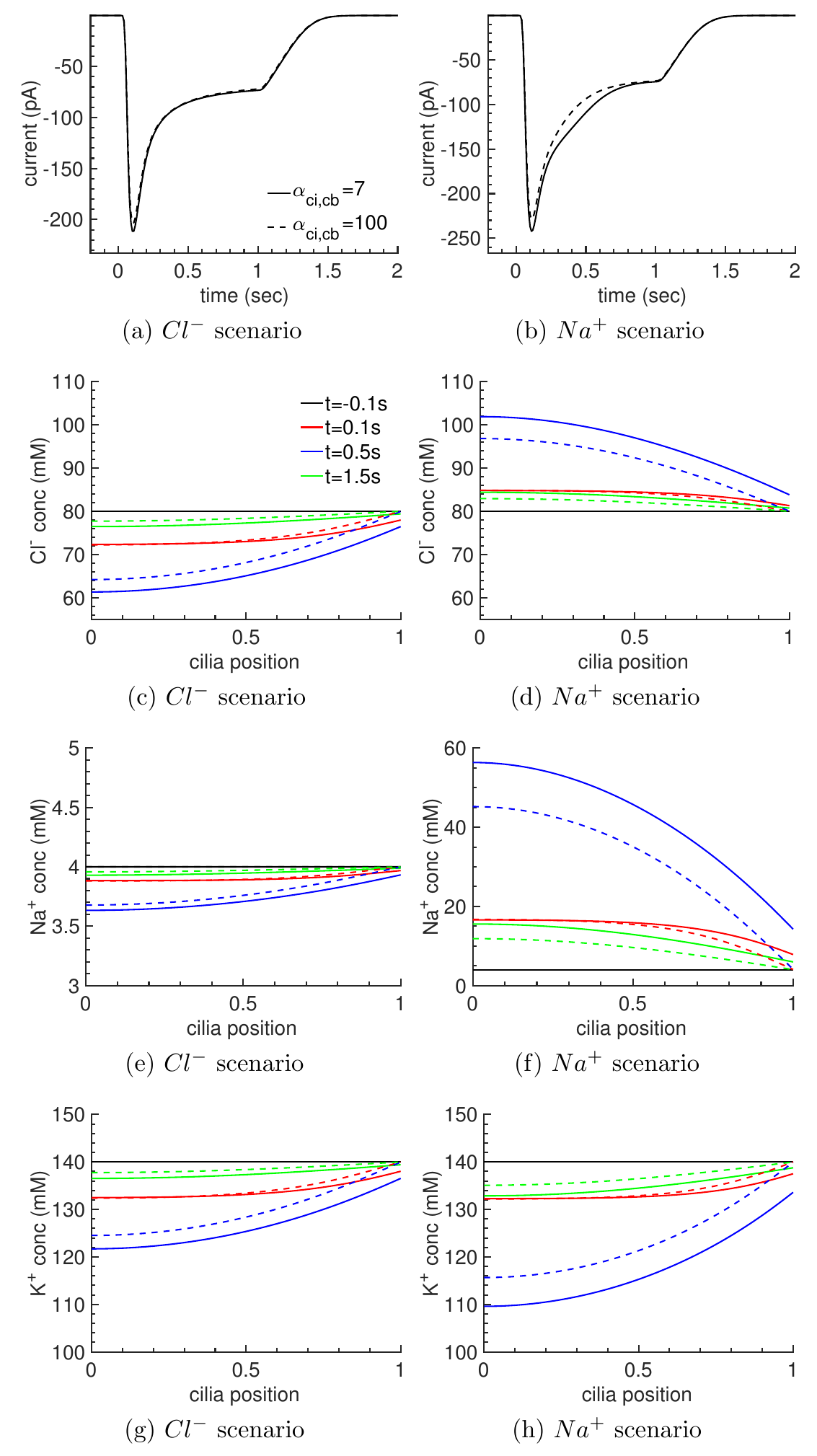}
		\caption{{\bf Flux vs Dirichlet boundary conditions.}  Comparison of simulations results with $\alpha_{ci,cb}=7$ (continuous lines) and $\alpha_{ci,cb}=100$ (dashed lines ) performed with the spatial model and strongest odorant stimulation. $\alpha_{ci,cb}=7$ is the default value used for simulations. A value $\alpha_{ci,cb}=100$ effectively corresponds to Dirichlet boundary conditions. } 
		\label{FigAlphaCiCb} 
	\end{center}
\end{figure}

In Fig.~\ref{FigAlphaCiCb} we compare simulation results obtained with $\alpha_{ci,cb}=7$ (continuous lines) to results obtained with $\alpha_{ci,cb}=100$ (dashed lines). $\alpha_{ci,cb}=7$ is the default value that we used for simulations, and $\alpha_{ci,cb}=100$ basically corresponds to Dirichlet boundary conditions (Fig.~\ref{FigAlphaCiCb}c-h). We find that the currents are only marginally affected by changing the boundary conditions (Fig.~\ref{FigAlphaCiCb}a-b). Dirichlet conditions slightly reduce the maximal fluctuations near the tip of a cilium, however, the main differences as discussed in the main text between $Cl^-$ and $Na^+$ scenario persist (Fig.~\ref{FigAlphaCiCb}c-f). Thus, our conclusions from the main text are robust against reasonable changes in the boundary conditions. 

Finally, in Fig.~\ref{FigMucus} we compare simulation results for the current obtained with the well-stirred model and mucosal concentrations of $Na^+$ and $Cl^-$ that are reduced from 140 to 70 and 40 mM. Contrary to Fig. 5 in the main text, with the well-stirred model we do not observe bistability. This finding suggests that diffusion and concentration gradients play a crucial role for bistability. It is beyond the scope of this work to perform a detailed analysis of this phenomena. A comprehensive analysis would have to determine not only the parameter space where two fixed points exist, but also their stability and the position of the boundary (separatrix) that separates their basin of attraction. This task is highly complex due to the non-linear equations, diffusional properties, exchanger and channel characteristics, and biochemistry that affects the opening of CNG channels, and beyond the scope of this work. 

\begin{figure}[htb]
	\begin{center}	
		\includegraphics[width=0.8\linewidth]{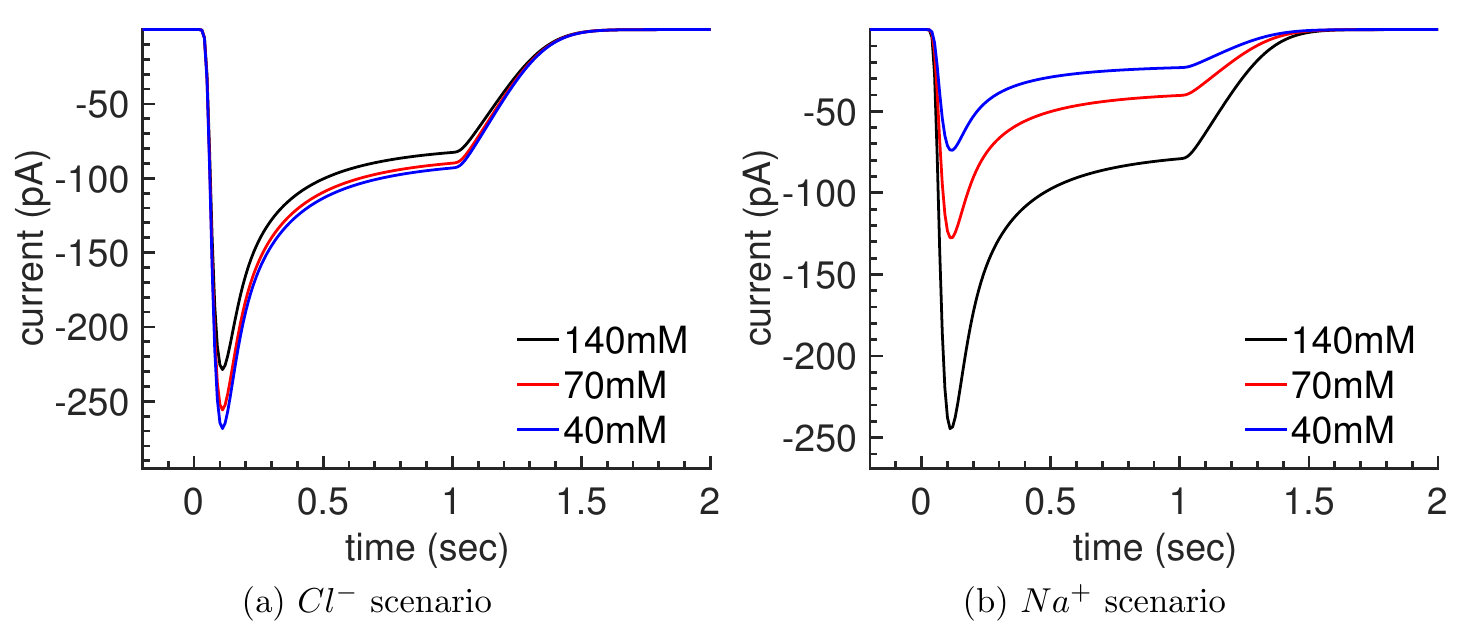}
		\caption{{\bf Well-stirred simulations with reduced mucosal concentrations for $Na^+$ and $Cl^-$. } Simulations are performed with the well-stirred model and strongest odorant stimulation (100 $\mu$M). Parameters are the same as in in Fig~\ref{FigFitting}, except for the mucosal $Na^+$ and $Cl^-$ concentrations that are indicated in the figure legend. Compared to the corresponding spatially resolved simulations in Fig. 5 in the main text, no bistability is observed with the well-stirred model.} 
		\label{FigMucus} 
	\end{center}
\end{figure}

\cleardoublepage

%%%%%%%%%%%%%%%%%%%%%%%%%%%%%%%%%%%%%%%%%%%%%%%%%%%%%%%%
\section{Parameters used for the simulations}
%%%%%%%%%%%%%%%%%%%%%%%%%%%%%%%%%%%%%%%%%%%%%%%%%%%%%%%%

\begin{table}[htbp]
	\begin{center}
		\begin{tabular}{|l|l|l|l|}
			\hline
			\textbf{Description} & \textbf{Symbol}  & \textbf{Unit } & \textbf{Value }\\
			\hline
			Cilium length  &   $L_{ci}$ & $\mu m$ & 25 \cite{Kleene_Review2008} \\
			Cilium radius &   $R_{ci}$ &  $\mu m$ & 0.075  \cite{Kleene_Review2008} \\
			Cilia number & $N_{ci}$ &  & 15  \cite{Kleene_Review2008}\\
			Diffusion constants (Ca$^{2+}$,Cl$^{-}$,Na$^+$,K$^+$) &  $D_{s}$  & $\mu m^2/s$ & 220 \cite{Reisertetal_ClCurrent_JGenPhys2003}, 2030, 1330, 1960  \cite{BookHille} \\	
			Mucus conc.  (Ringer) (Ca$^{2+}$,Cl$^{-}$,Na$^+$,K$^+$)    & $c^{s}_{mu}$   & mM & 2, 140, 140, 5 \\
			Cell body conc.    (Ca$^{2+}$,Cl$^{-}$,Na$^+$,K$^+$)    &  $c^{s}_{cb}$   & mM &  40 $\times 10^{-6}$, 80, 4, 140 \\
			Ano2 Ca$^{2+}$  sensitivity & $K_{ano}$ & $\mu M$ & 1.8 \cite{StephanReisertetal_Ano2_PNAS2009} \\
			Ano2 Ca$^{2+}$  Hill exponent & $h_{ano}$ & & 2.3 \cite{StephanReisertetal_Ano2_PNAS2009}\\
			Ano2 $Cl^-$ permeability  in the $Cl^-$ scenario & $\nu^{cl}_{ano}$ & $s^{-1}$ & 7.6\\
			Ano2 $Na^+$  permeability in the $Na^-$ scenario & $\nu^{cl}_{ano}$ & $s^{-1}$ & 3.4\\
			CNG cAMP sensitivity  & $K^{min}_{cng}$ & $\mu M$ & 4 \cite{Fringsetal_CNG_GenPhys1992,Songetal_Neuron2008}\\
			CNG cAMP Hill exponent & $h_{cng}$ & &  1.8 \cite{Fringsetal_CNG_GenPhys1992}\\
			CNG  $Ca^{2+}$   permeability & $\nu^{ca}_{cng}$ & $s^{-1}$ & 0.5 \\
			CNG  $Na^{+}$ permeability & $\nu^{na}_{cng}$ & $s^{-1}$ & 0 \\
			CNG  $K^{+}$ permeability   & $\nu^{k}_{cng}$ & $s^{-1}$ & 0\\
			CNG Ca$^{2+}$ feedback sensitivity  & $K^{ca}_{cng}$ & $\mu M$ & 10  \cite{ChenYau_CNG_Nature1994} \\
			CNG Ca$^{2+}$ feedback  max & $f^{ca}_{cng}$ &  & 4 \cite{ChenYau_CNG_Nature1994} \\
			NCKX4 rate constant & $\nu_{nckx}$ & $s^{-1}$  & 1.2\\
			NCKX4 Michaelis constant  & $K_{nckx}$ & $\mu M$ & 22\\
			Rate constant for efflux from a cilium & $\alpha_{ci,cb}$ & $s^{-1}$ & 7 \\
			Cell body leak voltage & $U_{cb}^{leak}$ & mV& -65\\
			Cell body leak conductance & $g_{cb}^{leak}$ & nS & 20 \\
			Cell body capacitance & $C_{cb}$ & nF& $10^{-3}$ \\	
			Ciliary membrane capacity  & $C_m$  & ${nF}/{\mu m^2}$ & $10^{-5}$\\
			Fast buffer constant   &  $B_{ca}$   &  &  0 \\
			Slow buffer total concentration   &  $c^{cab}_{ci,tot}$   & mM &  0 \\	
			\hline
		\end{tabular}
		\caption{Parameters for electrodiffusion. }
		\label{table1}
	\end{center}
\end{table}

\begin{table}[htbp]
	\begin{center}
		\begin{tabular}{|l|l|l|l|}
			\hline
			\textbf{Description} & \textbf{Symbol}  & \textbf{Unit } & \textbf{Value }\\
			\hline
			cAMP diffusion constant  & $D_{camp}$ & $\mu m^2/s$  & 270 \cite{CygnarZhao_PDE_NatNEurosc2009,ChenEtal_BiophysJ1999}\\
			cAMP maximal synthesis rate  & $\alpha^{max}_{camp}$ & $mu M s^{-1}$  &  95\\
			cAMP hydrolysis rate  & $\beta_{camp}$ & $s^{-1}$  &  50\\
			cAMP cell body concentration  & $c^{camp}_{cb}$ & $\mu M$ & 0\\
			CaMK time constant  &$beta_{camk}$  & $s^{-1}$   & 0.7\\
			CaMK max inhibition  &$f^{max}_{camk}$  &    & 28\\
			CaMK $Ca^{2+}$ sensitivity  &$K_{camk}$  &  $\mu M$   & 2\\
			CaMK $Ca^{2+}$ Hill exp  &$h_{camk}$  &     & 3\\
			Receptor sensitivity   &$K_{od}$  &  $\mu M$   & 45\\
			Receptor Hill exponent   &$h_{od}$  &    & 2\\
			G protein sensitivity   &$K_{or}$  &    & 0.7\\
			G protein time scale   &$\beta_{g}$  &  $s^{-1}$   & 6.4\\
			AC time scale   &$\beta_{ac}$  &  $s^{-1}$   & 20\\
			AC sensitivity     &$K_{g}$  &    & 0.1 \\
			\hline
		\end{tabular}
		\caption{Parameters for the biochemical transduction model. }
		\label{table2}
	\end{center}
\end{table}

\end{appendix}

\cleardoublepage

% Bibliography

%\bibliographystyle{ieeetr}
%\bibliography{/Users/reingrub/Documents/Science/MyProjects/bibliography/Bib_Olfaction,/Users/reingrub/Documents/Science/MyProjects/bibliography/Bib_ReactionTheoryDiffusionNET,/Users/reingrub/Documents/Science/MyProjects/bibliography/Bib_Books,/Users/reingrub/Documents/Science/MyProjects/bibliography/Bib_Phototransduction,/Users/reingrub/Documents/Science/MyProjects/bibliography/Bib_Holcman,/Users/reingrub/Documents/Science/MyProjects/bibliography/Bib_Reingruber}

\end{document}